\documentclass[12pt,journal,draftcls,letterpaper,onecolumn]{IEEEtran}
\usepackage{epsfig}
\usepackage{graphicx}
\usepackage{amsmath}
\usepackage{amssymb}
\usepackage{amsfonts}
\usepackage{algorithm}
\usepackage{algorithmic}
\usepackage{rotating}
\usepackage{subfigure}
\usepackage{multirow}
\usepackage{multicol}
\usepackage{blindtext}
\usepackage{float}
\newtheorem{lemma}{Lemma}
\newtheorem{theorem}{Theorem}
\newtheorem{definition}{Definition}
\newtheorem{proposition}{Proposition}
\newtheorem{corollary}{Corollary}
\begin{document}

\title{Hamming Compressed Sensing}

\author{Tianyi~Zhou,~
        and Dacheng~Tao,~\IEEEmembership{Member,~IEEE}
\thanks{}
\thanks{T. Zhou and D. Tao are with the Centre for Quantum Computation \& Intelligent Systems, University of Technology, Sydney, Australia, NSW 2007.}}


\maketitle

\begin{abstract}
Compressed sensing (CS) and 1-bit CS cannot directly recover quantized signals and require time consuming recovery. In this paper, we introduce \textit{Hamming compressed sensing} (HCS) that directly recovers a k-bit quantized signal of dimensional $n$ from its 1-bit measurements via invoking $n$ times of Kullback-Leibler divergence based nearest neighbor search. Compared with CS and 1-bit CS, HCS allows the signal to be dense, takes considerably less (linear) recovery time and requires substantially less measurements ($\mathcal O(\log n)$). Moreover, HCS recovery can accelerate the subsequent 1-bit CS dequantizer. We study a quantized recovery error bound of HCS for general signals and ``HCS+dequantizer'' recovery error bound for sparse signals. Extensive numerical simulations verify the appealing accuracy, robustness, efficiency and consistency of HCS.
\end{abstract}

\IEEEpeerreviewmaketitle

\section{Introduction}

\IEEEPARstart{D}{igital} revolution triggered a rapid growth of novel signal acquisition techniques with primary interests in reducing sampling costs and improving recovery efficiency. The theoretical promise of conventional sampling methods comes from the Shannon/Nyquist sampling theorem \cite{Shannon}, which states a signal can be fully recovered if it is sampled uniformly at a rate more than twice its bandwidth. Such uniform sampling is always done by analog-to-digital converters (ADCs). Unfortunately, for many real applications such as radar imaging and magnetic resonance imaging (MRI), Nyquist rate is too high due to the expensive cost of analog-to-digital (AD) conversion, the maximum sampling rate limits of the hardware, or the additional costly compression to the obtained samples.

\subsection{Compressed sensing}

Recently, prosperous researches in compressed sensing (CS) \cite{DonohoCS}\cite{CandesT06}\cite{ErrorCS}\cite{ModelCS}\cite{BayesianCS}\cite{SparseCS} show that an accurate recovery can be obtained by sampling signals at a rate proportional to their underlying ``information content'' rather than their bandwidth. The key improvement brought by CS is that the sampling rate can be significantly reduced by replacing the uniform sampling with linear measurement, if the signals are sparse or compressible on certain dictionary. This improvement leverages the fact that many signals of interest occupies a quite large bandwidth but has a sparse spectrum, which leads to a redundancy of the uniform sampling. In particular, CS dedicates to reconstruct a signal $x\in\mathbb R^n$ from its linear measurements
\begin{equation}
\notag y=\Phi x=\Phi\Psi\alpha,
\end{equation}
where $\Phi\in\mathbb R^{m\times n}$ is the measurement or sensing matrix allowing $m\ll n$ (in which case $\Phi$ is an underdetermined system). The signal $x$ is $K$-sparse if its nonzero entries are less than $K$. Given a dictionary $\Psi$, $x$ is $K$-compressible if the nonzero entries of $\alpha$ are less than $K$. If sparse/compressible $x$ is the Nyquist-rate samples of a analog signal $x(t)$, CS replaces ADCs with a novel sampler $\bar\Phi$ such that $y=\Phi x=\bar\Phi(x(t))$. A straightforward approach for recovering $x$ from $y$ is to minimize the number of nonzero entries in $x$, i.e., the $\ell_0$ norm of $x$. Specifically, it is not difficult to demonstrate that a $K$-sparse signal $x$ can be accurately recovered from $2K$ measurements by solving
\begin{equation}
\notag \min_{x\in\mathbb R^n} \left\|x\right\|_0~~s.t.~~y=\Phi x~~~~~~{\rm(CS_{\ell_0})}
\end{equation}
with exhaustive search if $\Phi$ is a generic matrix. However, such exhaustive search has intractable combinatorial complexity. So some CS methods adopts iterative and greedy algorithms to solve the $\ell_0$ minimization, such as orthogonal matching pursuit (OMP) \cite{OMP}, compressive sampling matching pursuit (CoSaMP) \cite{CoSaMP}, approximate message passing \cite{MessagePassing}, iterative splitting and thresholding (IST) \cite{IST} and iterated hard shrinkage (IHT) \cite{IterativeHardThresh}.

Since $\ell_0$ minimization is a non-convex problem, some other CS methods solve its convex relaxation, i.e., $\ell_1$ minimization and its variants:
\begin{equation}
\notag \min_{x\in\mathbb R^n} \left\|x\right\|_1~~s.t.~~y=\Phi x.~~~~~~{\rm(CS_{\ell_1})}
\end{equation}

Various convex optimization approaches have been developed or introduced to solve the above problem and its variants. Representatives include basis pursuit \cite{BasisPursuit}, Dantzig selecter \cite{DantzigSelector}, NESTA \cite{NESTA}, interior point method \cite{InteriorPoint}, coordinate gradient descent \cite{CoordinateDescent}, gradient projection \cite{GPSR} and the class of approaches based on the fixed point method such as Bregman iterative algorithm \cite{BregmanIteration}, fixed point continuation \cite{FPC} and iteratively re-weighted least squares (IRLS) \cite{IRLS}. It is also worthy noting that the \emph{lasso} \cite{lasso} type algorithms \cite{LARS}\cite{GroupLasso} for model selection can be applied to CS recovery problem as well.

However, compared with the recovery schemes in conventional sampling theory, which reconstructs signals by invoking nearly linear transformation, most aforementioned recovery algorithms in CS methods require polynomial time cost, which is substantially more expensive than the conventional methods. This burden in recovery efficiency limits the applications of CS in many real problems, in which the dimension of the signals is extremely high.

Beyond the recovery efficiency, another important issue in CS is the theoretical guarantee for precise recovery. Since most existing CS algorithm finds the signal that agrees with the measurements $y$ without directly minimizing $\ell_0$ norm, the recovery success of existing CS methods relies on another theoretical requirement for $\Phi$ (or $\Phi\Psi$ in compressible case), i.e., two sufficiently close measurements $y_1=\Phi x_1$ and $y_2=\Phi x_2$ indicates that vectors $x_1$ and $x_2$ are sufficiently close to each other. This low-distortion property of the linear operator $\Phi$ is called ``restricted isometry property (RIP)'' \cite{CandesRT06}\cite{StableCS}, which can also be interpreted as the incoherence between the measurement matrix $\Phi$ and the signal dictionary $\Psi$ (or identity matrix $I$ for sparse $x$) in order to restrict the concentration of a single $\alpha_i$ (or $x_i$ in sparse case) in the measurements. An intriguing property of CS is that some randomly generated matrices such as Gaussian, Bernoulli and partial Fourier ensemble fulfill RIP with high probabilities. For example, $\Phi$ whose entries are randomly drawn from a sub-Gaussian distribution satisfies RIP with a high probability if $m=\mathcal O(K\log(n/K))$. By using the concept of RIP, the global solution of the $\ell_1$ minimization with such $\Phi$ is guaranteed to be sufficiently close to the original sparse signal. Thus CS can successfully recover $K$-sparse signals of dimension $n$ from $m=\mathcal O(K\log(n/K))$ measurements. However, given a deterministic $\Phi$ with $m=\mathcal O(K\log(n/K))$, it is generally regarded as NP-hard to test whether RIP holds or not.

In practice, it is paramount that signals are not exactly sparse and the measurements cannot be precise due to the hardware limits. Questions arise in this case that ``is it possible to recovery the $K$ largest entries if $x$ is nearly sparse?'' and ``is it possible to recover $x$ from noisy measurements $y$?''. These questions lead to the problem of stable recovery \cite{StableCS}\cite{EldarRobust} in CS. Fortunately RIP can naturally address this problem, because it ensures that small changes in measurements induce small changes in the recoveries. In stable recovery, the constraint $y=\Phi x$ in the original $\ell_0$ and $\ell_1$ minimization problems are replaced with $\left\|y-\Phi x\right\|_2\leq\epsilon$. Another variant in this case is minimizing $\left\|y-\Phi x\right\|_2$ with penalty or constraint to $\ell_0$ or $\ell_1$ norm of $x$. Many existing CS algorithms such as basis pursuit denoising (BPDN) \cite{BasisPursuit} can also handle the stable recovery problem.

Today's state-of-the-art researches in CS focus primarily on further reducing the number of measurements, improving the recovery efficiency and increasing the robustness of stable recovery. Although CS \cite{CSreview} exhibits powerful potential in simplifying the sampling process and reducing the measurement amount, there are some unsettled issues of CS when applied to realistic digital systems, especially in developing hardware. A crucial problem is how to deal with the quantization of the measurements.

\subsection{Quantized compressed sensing}

In practical digital systems, quantization of CS measurements $y$ is a natural and inevitable process, in which each measurement is transformed from a real value to a finite number of bits that represent a finite interval containing the real value. In CS, quantization is an irreversible process introducing error in measurements, if we round the quantized measurement as any real value within the corresponding interval in recovery. One commonly used trick to deal with quantization error is to treat it as bounded Gaussian noise in measurements, and thus stable recovery methods in CS guarantee to obtain a robust recovery. However, this solution cannot produce acceptable recovery result unless the quantization error is sufficiently small and near Gaussian. Unfortunately, these two conditions are often hardly fulfilled because 1) small quantization error is obtained from small interval width, which is the result of high sampling rate and high conversion accuracy of ADC. This conflicts with the spirit of CS; and 2) the quantization error is usually highly non-Gaussian.

Several recent works \cite{BPDNQ}\cite{QCSboyd}\cite{1bitCS}\cite{Distorsion} address the quantized compressed sensing (QCS) by (implicitly or explicitly) treating the quantization as a box constraint to the measurements $y=\Phi x+\epsilon$,
\begin{equation}
\notag \min_{x\in\mathbb R^n} \left\|x\right\|_1~~s.t.~~u\leq\Phi x+\epsilon\leq v, ~~~~~~{\rm(QCS)}
\end{equation}
where the two vectors $u$ and $v$ store the corresponding boundaries of the intervals that the entries of $y$ lie in, and $\epsilon$ is the measurement noise. The box constraint is also called quantization consistency (QC) constraint. By solving this problem, the quantization error will not be wholly transformed into the recovery error via RIP. Thus it is possible to obtain an accurate recovery from very coarsely quantized measurements. A variant of BPDN called ``basis pursuit dequantizer'' proposed in \cite{BPDNQ} restricts $\left\|y-\Phi x\right\|_p$($2\ll p\ll\infty$) rather than $\left\|y-\Phi x\right\|_2$ in $\ell_1$ norm minimization, and proves that the recovery error decreases by a factor of $\sqrt{p+1}$. In \cite{Distorsion}, an adaption of BPDN and subspace pursuit \cite{SubspacePursuit} integrate an explicit QC constraint. An $\ell_1$ regularized maximum likelihood estimation is developed in \cite{QCSboyd} to solve QCS with noise (i.e., $\epsilon\neq 0$).

As the extreme case of QCS, 1-bit CS \cite{1bitCS}\cite{Robust1bitCS} has been developed to reconstruct sparse signals from 1-bit measurements, which merely capture the signs of the linear measurements in CS. The 1-bit measurements enable simple and fast quantization. Thus it can significantly reduce the sampling costs and strengthen the robustness of hardware implementation. One inevitable information loss in 1-bit measurements is the scale of the original signal, because scaled signal will have the same linear measurement signs as the original one. Theoretically, 1-bit CS ensures consistent reconstructions of signals on the unit $\ell_2$ sphere \cite{Greedy1bitCS}\cite{Trust}. In 1-bit CS, one-sided $\ell_2$ \cite{1bitCS} or $\ell_1$ \cite{Robust1bitCS} objectives are designed to guarantee the consistency of 1-bit measurements by imposing a sign constraint or minimizing the sign violations in optimization. Analogous to RIP in CS, the binary $\epsilon$-stable embedding (B$\epsilon$SE) \cite{Robust1bitCS} ensures the low distortion between the original signals and their 1-bit measurements, and thus guarantees the accuracy and stableness of the reconstruction. It is remarkable that $m=\mathcal O(K\log n)$ can guarantees B$\epsilon$SE and the subsequent successful recovery. Most 1-bit CS recovery algorithms, e.g., renormalized fixed point iteration \cite{1bitCS}, matched sign pursuit \cite{Greedy1bitCS} and binary iterative hard thresholding (BIHT) \cite{Robust1bitCS}, are extensions of CS recovery algorithms. It has been shown that BIHT, a variant of IHT \cite{IterativeHardThresh}, can produce precise and consistent recovery from 1-bit measurements.

QCS and 1-bit CS not only consider the quantization of the measurements but also improve the recovery robustness to the nonlinear distortions brought by ADC, because the quantized measurements only preserve the intervals the real-value measurements lie in. However, QCS and 1-bit CS methods require polynomial-time recovery algorithms and thus they are prohibitive to high dimensional signals in practical applications. Moreover, another central problem is that either CS or QCS recovers the original real-value signals, but quantization of the recovered signals is inevitable in digital systems.

\subsection{Hamming compressed sensing}

Digital systems prefer to use the quantized recovery of the original signal, which can be processed directly, but the recoveries of both CS and QCS are continuous and real-valued. In order to apply them to digital systems, a straightforward solution is to impose an additional quantization to the CS or QCS recoveries. However, this quantization requests additional time costs and expenses on ADCs, which could be expensive if the sampling rate is required to be high. Moreover, the convex optimization or iterative greedy search based recovery in CS and QCS is of polynomial-time. This is not acceptable for high-dimensional signals. In addition, the trade-off between the recovery time and the recovery resolution cannot be controlled in CS and QCS, although it is preferred in practice. Finally, the success of CS and QCS is based upon the assumption that signals are sparse. When the signal $x$ is dense, the numbers of measurements are large required by CS and QCS and the advantages of CS and QCS are lost accordingly.

In this paper, we directly recover the quantization of a general signal (not necessary to be $K$-sparse) from the quantization of its linear measurements, or ``quantized recovery (QR)''. In particular, for a signal $x$ and its quantization $q=Q(x)$ by a quantizer $Q(\cdot)$, we seek for a recovery algorithm $R(\cdot)$ that reconstructs $q^*=R(y)$ sufficiently close to $q$ from the quantized measurements $y=A(x)$, where the operator $A$ is a composition of linear measurement and quantization. This problem has not been formally studied before, and has the potential to mitigate the aforementioned limitations of CS and QCS. The main motivation behind QR is sacrificing the quantization error of the recovery for reducing the number of measurements. Thus the recovery time can be significantly reduced with the decreasing of the number of bits for the quantized recovery, and the number of measurements can be small even when the signal is dense. Comparing with CS and QCS, QR considers the quantization error of the quantized recovery in determining the sampling rate and developing the reconstruction algorithm.

The primary contribution of this paper is developing \textit{Hamming compressed sensing} (HCS) to achieve quantized recovery from a small number of quantized measurements with extremely small time cost and without signal sparsity constraint. In compression (sampling), we adopt the 1-bit measurements \cite{Robust1bitCS} to guarantee consistency and B$\epsilon$SE but employ them in a different way. In particular, we introduce a bijection between each dimension of the signal and a Bernoulli distribution. The underlying idea of HCS is to estimate the Bernoulli distribution for each dimension from the 1-bit measurements, and thus each dimension of the signal can be recovered from the corresponding Bernoulli distribution. In order to define the quantized recovery, we propose a k-bit HCS quantizer splitting the signal domain into $k$ intervals, which are derived from the bijection as the mappings of the $k$ uniform linear quantization boundaries for the Bernoulli distribution domain. In recovery, HCS searches the nearest neighbor of the estimated Bernoulli distribution among the $k$ boundaries in the Bernoulli distribution domain, and recovers the quantization of the corresponding dimension as the HCS quantizer interval associated with the nearest boundary. We theoretically study a quantized recovery error bound of HCS by investigating the precision of the estimation and its impact on the KL divergence based nearest neighbor search. The theoretical analysis provide a strong support to the successful recovery of HCS.

Comparing with CS and QCS, HCS has the following significant and appealing merits:

\begin{itemize}

\item[1] HCS provides simple and low-cost sampling and recovery schemes for digital systems. The procedures are substantially simple: the sampling and sensing are integrated to 1-bit measurements, while the recovery and quantization are integrated to quantized recovery. Furthermore, both the 1-bit measurement and the quantized recovery do not require ADC with a high sampling rate. Note that HCS remains the recovery robustness due to quantized measurements inherited from QCS.

\item[2] The recovery in HCS only requires to compute $nk$ Kullback-Leibler (KL) divergences for obtaining k-bit recovery of an $n$-dimensional signal, and thus is a non-iterative, linear algorithm. The recovery includes very simple computations. Therefore, HCS is considerably more efficient and easier to be implemented than CS and QCS.

\item[3] According to the theoretical analysis of HCS, merely $m=\mathcal O(\log n)$ 1-bit measurements are sufficient to produce a successful quantized recovery with high probability. Note there is no sparse assumption to the signal $x$. Therefore, HCS allows more economical compression than CS and QCS.

\end{itemize}

Another compelling advantage of HCS is it can promote the recovery of the real-value signals after quantized recovery. When the subsequent dequantization $x^*=D(q^*)$ after quantized recovery is required, we can treat the HCS quantized recovery as a box constraint to reduce the search space of the 1-bit CS dequantizer $D(\cdot)$ in order to accelerate the convergence. By invoking the HCS recovery bound, the consistency and B$\epsilon$SE from 1-bit CS, we show an error bound of ``HCS+dequantizer'' recovery for sparse signals.

The rest of this paper is outlined as follows. Section 2 introduces the 1-bit measurements in HCS, which lead to a bijection between each dimension of the signal and a Bernoulli distribution, and its consistency. Section 3 presents the k-bit reconstruction in HCS, including how to obtain the HCS quantizer, KL-divergence nearest neighbor search based recovery and theoretical evidence for successful recovery. Section 4 introduces the application of HCS recovery results in dequantization, an theoretical analysis of the dequantization error is given here. Section 5 shows the power of HCS via three groups of experiments. Section 6 concludes.

\section{1-bit Measurements}

HCS recovers the quantized signal directly from its quantized measurements, each of which is composed of a finite number of bits. We consider the extreme case of 1-bit measurements of a signal $x\in\mathbb R^n$, which are given by
\begin{equation}\label{E:1bitm}
y=A(x)={\rm sign}\left(\Phi x\right),
\end{equation}
where ${\rm sign}(\cdot)$ is an element-wise sign operator and $A(\cdot)$ maps $x$ from $\mathbb R^n$ to the Boolean cube $\mathbb B^M:=\{-1, 1\}^M$. Since the scale of the signal is lost in 1-bit measurements $y$ (multiplying $x$ with a positive scalar will not change the signs of the measurements), the consistent reconstruction can be obtained by enforcing the signal $x\in\Sigma^*_K:=\{x\in S^{n-1}:\|x\|_0\leq K\}$ where $S^{n-1}:=\{x\in\mathbb R^n:\|x\|_2=1\}$ is the $n$-dimensional unit hyper-sphere.

The 1-bit measurements $y$ can also be viewed as a ``hash'' of the signal $x$. Similar hash based on random projection signs is developed in locality sensitive hashing (LSH) \cite{LSH}\cite{LSH2008}. LSH performs an approximate nearest neighbor (ANN) searches on the hashes of signals, and proves the results approach the precise NN searches on the original signals with high probability. This theoretical guarantee is based on condition similar to B$\epsilon$SE \cite{Robust1bitCS} in 1-bit CS. It is interesting to compare LSH with HCS, because LSH is an irreversible process aiming at ANN, while HCS can be viewed as a reversible LSH in this case.


\subsection{Bijection}

In contrast to CS and 1-bit CS, HCS does not recover the original signal, but reconstructs the quantized signal by recovering each dimension in isolation. In particular, according to Lemma 3.2 in \cite{AppMaxCut}, we show that there exists a bijection (cf. Theorem \ref{T:PrRPSign}) between each dimension of the signal $x$ and a Bernoulli distribution, which can be uniquely estimated from the 1-bit measurements. The underlying idea of HCS is to estimate the Bernoulli distribution for each dimension, and recover the quantization of the corresponding dimension as the interval where the Bernoulli distribution's mapping lies in.

\begin{theorem}\label{T:PrRPSign}
{\rm(\textbf{Bijection})} For a normalized signal $x\in\mathbb R^n$ with $\|x\|_2=1$ and a normalized Gaussian random vector $\phi$ that is drawn uniformly from the unit $\ell_2$ sphere in $\mathbb R^n$ (i.e., each element of $\phi$ is firstly drawn i.i.d. from the standard Gaussian distribution $\mathcal N(0,1)$ and then $\phi$ is normalized as $\phi/\|\phi\|_2$), given the $i^{th}$ dimension of the signal $x_i$ and the corresponding coordinate unit vector $e_i=\{0,\cdots,0,1,0,\cdots,0\}$, where $1$ appears in the $i^{th}$ dimension, there exists a bijection $P:\mathbb R\rightarrow\mathbb P$ from $x_i$ to the Bernoulli distribution of the binary random variable $s_i={\rm sign}\left(\langle x,\phi\rangle\right)\cdot{\rm sign}\left(\langle e_i,\phi\rangle\right)$:
\begin{equation}\label{E:PrRPSign}
P(x_i)=\left\{
  \begin{array}{ll}
    \Pr\left(s_i=-1\right)=\frac{1}{\pi}\arccos\left(x_i\right), \\
    \Pr\left(s_i=1\right)=1-\frac{1}{\pi}\arccos\left(x_i\right).
  \end{array}
\right.
\end{equation}
\end{theorem}

Since the mapping between $x_i$ and $P(x_i)$ is bijective, given $P(x_i)$, the $i^{th}$ dimension of $x$ can be uniquely identified. According to the definition of $s_i$, $P(x_i)$ can be estimated from the instances of the random variable ${\rm sign}\left(\langle x,\phi\rangle\right)$, which are exactly the 1-bit measurements $y$ defined in (\ref{E:1bitm}). Therefore, the 1-bit measurements $y$ include sufficient information to reconstruct $x_i$ from the estimation of $P(x_i)$, and the recovery accuracy of $x_i$ depends on the accuracy of the estimation to $P(x_i)$.

\subsection{Consistency}

Given a signal $x$, its quantization $q=Q(x)$ by HCS quantizer $Q(\cdot)$, the quantized recovery $q^*=R(y)$ obtained by HCS reconstruction $R(\cdot)$ from the 1-bit measurements $y=A(x)$, and its dequantization $x^*=D(q^*)$ obtained by a dequantizer $D(\cdot)$, the ``HCS+dequantizer'' recovery $x^*$ is given by
\begin{equation}\label{E:err}
x^*=x+err_H+err_D,
\end{equation}
where $err_H$ is determined by the difference between $q$ and $q^*$ caused by HCS reconstruction, and $err_D$ is the dequantization error from $q^*$ to $x^*$. The upper bounds of $err_H$ and $err_D$ will be given in Sections 4 and 5, respectively. The following Lemma \ref{L:Consis} shows the consistency pertaining to $err_H$ and $err_D$.

\begin{lemma}\label{L:Consis}
{\rm(\textbf{Consistency})} Let $\Phi$ be a standard Gaussian random matrix whose $m$ rows are composed of $\phi_i$ defined in Theorem \ref{T:PrRPSign}. The measurement operator $A(\cdot)$ is defined in (\ref{E:1bitm}). Given a fixed $\gamma>0$, for any signal $x\in\mathbb R^n$ and its ``HCS+dequantizer'' recovery $x^*$, we have
\begin{align}
&\mathbb E\left(D_H\left(A(x^*),A(x)\right)\right)\leq g(\sigma,\|x\|_2),\\
&\Pr\left(D_H\left(A(x^*),A(x)\right)>g(\sigma,\|x\|_2)+\gamma\right)\leq e^{-2m\gamma^2},
\end{align}
where $D_H(u,v)=\frac{1}{M}\sum_{i=1}^{M}u_i\oplus v_i$ ($u,v\in\mathbb\{-1,1\}^M$) is the normalized Hamming distance, $g(\sigma,\|x\|_2)=\frac{1}{2}\frac{\sigma}{\sqrt{\|x\|_2^2+\sigma^2}}\leq \frac{1}{2}\frac{\sigma}{\|x\|_2}$ and $\sigma=\left\|err_H+err_D\right\|_2$.
\end{lemma}
\begin{proof}
According to (\ref{E:1bitm}) and (\ref{E:err}), we have
\begin{equation}
A(x^*)={\rm sign}\left(\Phi x^*\right)={\rm sign}\left(\Phi x+\Phi\left(err_H+err_D\right)\right).
\end{equation}
Since $err=\Phi\left(err_H+err_D\right)$ is a Gaussian random noise vector whose $i^{th}$ element $err_i\sim\mathcal N(0,\sigma^2)$. According to Lemma 5 in \cite{Robust1bitCS}, we obtain Lemma \ref{L:Consis}. This completes the proof.
\end{proof}

The consistent reconstruction in CS and 1-bit CS minimizes $\left\|A(x)-A(x^*)\right\|$ for a $K$-sparse signal $x$. RIP and B$\epsilon$SE bridge the consistency and the reconstruction accuracy in CS and 1-bit CS, respectively. Instead of minimizing $\left\|A(x)-A(x^*)\right\|$ to achieve the recovery accuracy, HCS directly estimates the interval that each dimension of the signal $x$ lies in from the estimated Bernoulli distribution defined in Theorem \ref{T:PrRPSign}.

In addition, the consistency between $A(x)$ and $A(x^*)$ is important for HCS, because in part it determines 1) the amount of information preserved in 1-bit measurements, and 2) the error bound of ``HCS+dequantizer'' recovery for sparse signals.

\section{k-bit Reconstruction}

The primary contribution of this paper is the quantized recovery in HCS, which reconstructs the quantized signal from its 1-bit measurements (\ref{E:1bitm}). Figure \ref{fig:HCSscheme} illustrates HCS quantized recovery. To define the HCS quantizer, we firstly find $k$ boundaries $P_j$($j=0,\cdots,k-1$) (\ref{E:Bernq}) in Bernoulli distribution domain by imposing the uniform linear quantizer to the range of $P_j^-$. Given an arbitrary $x_i$, the nearest neighbor of $P(x_i)$ among the $k$ boundaries $P_j$($j=0,\cdots,k-1$) indicates the interval $q_i$ that $x_i$ lies in the signal domain. The $k+1$ boundaries $S_j$($j=0,\cdots,k$) associated with the $k$ intervals $q_j$($j=0,\cdots,k$) are calculated from the $k$ boundaries $P_j$($j=0,\cdots,k-1$) according to the bijection defined in Theorem \ref{T:PrRPSign}. In HCS recovery, $P(x_i)$ is estimated as $\hat P(x_i)$ from the 1-bit measurements $y$. Then the nearest neighbor of $\hat P(x_i)$ among the $k$ boundaries $P_j$($j=0,\cdots,k-1$) is determined by comparing the KL-divergences between $\hat P(x_i)$ and $P_j$. The quantization of $x_i$ defined by HCS quantizer is recovered as the interval $q_i$ corresponding to the nearest neighbor.

In this section, we first introduce the HCS quantizer, which is a mapping resulting from the uniform linear quantizer of the Bernoulli distribution domain to the signal domain. The quantized recovery procedure is composed of $n$ times of KL-divergence based nearest neighbor searches. Thus it is a linear algorithm much faster than the conventional reconstruction algorithms of CS and 1-bit CS, which require optimization with the $\ell_p(0\leq p\leq2)$ constraint/penalty, or iterative thresholding/greedy search. We then study the upper bound of the quantized recovery error $err_H$.


\subsection{HCS quantizer}

Since HCS aims at recovering the quantization of the original signal, we firstly introduce HCS quantizer, which defines the intervals and boundaries for quantization in the signal domain. These intervals and boundaries are uniquely derived from a predefined uniform linear quantizer in the Bernoulli distribution domain. Given a signal and the boundaries of HCS quantizer, its k-bit HCS quantization can be identified. We will show HCS quantizer performs closely to the uniform linear quantizer.

Note that in the quantized recovery of HCS, the reconstruction and quantization are simultaneously accomplished. Thus the HCS quantizer will not play an explicit role in the recovery procedure. However, it is related to and uniquely determined by the quantization of the Bernoulli distribution domain, which plays an important role in the recovery and explains the reconstruction $q^*$. Moreover, it will be applied to the error bound analyses for $err_H$ and $err_H+err_D$.

\begin{figure*}[htb]
\begin{center}
\subfigure{
 \includegraphics[width=0.76\linewidth]{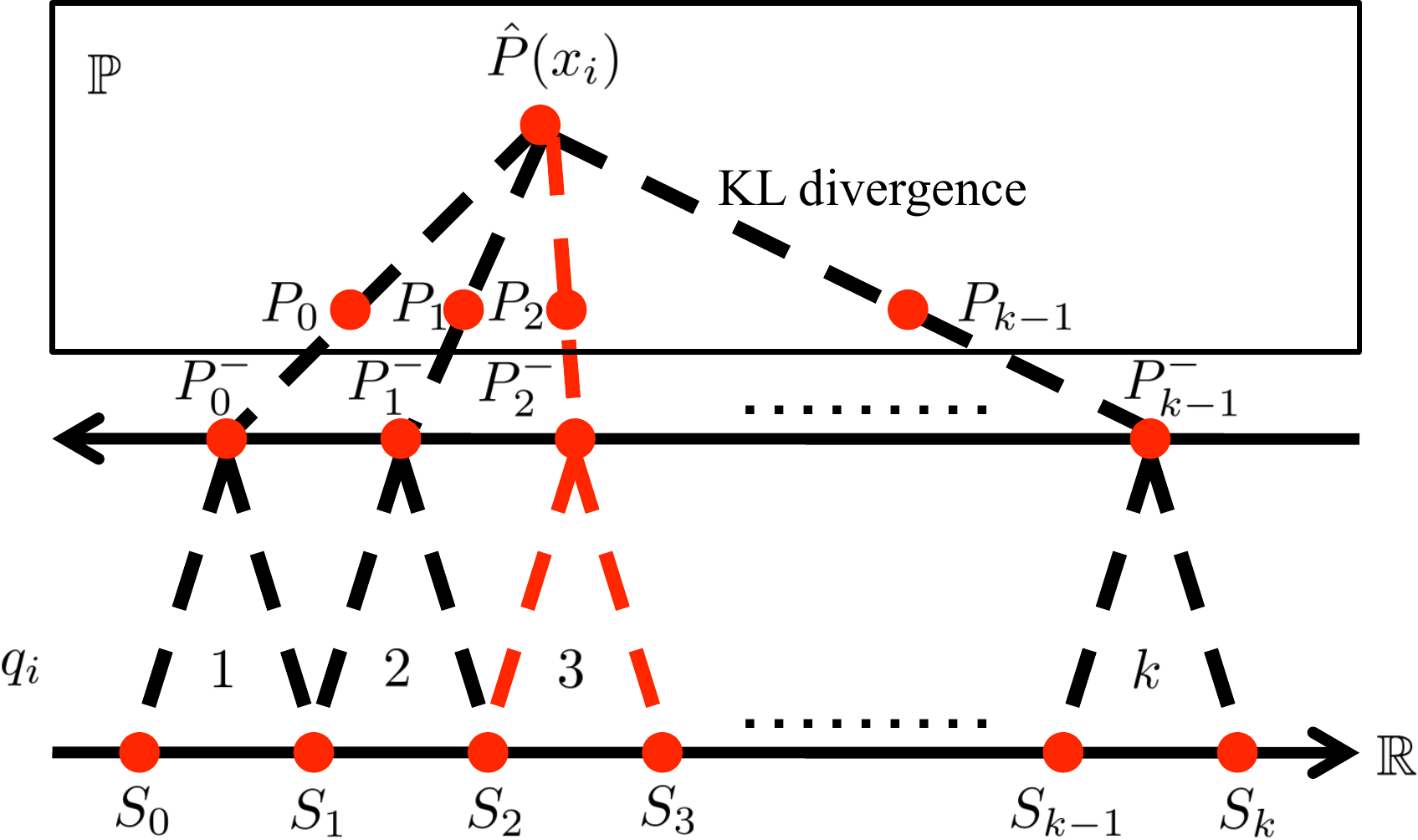}
 \label{fig:HCSscheme}
}
\subfigure{
 \includegraphics[width=0.88\linewidth]{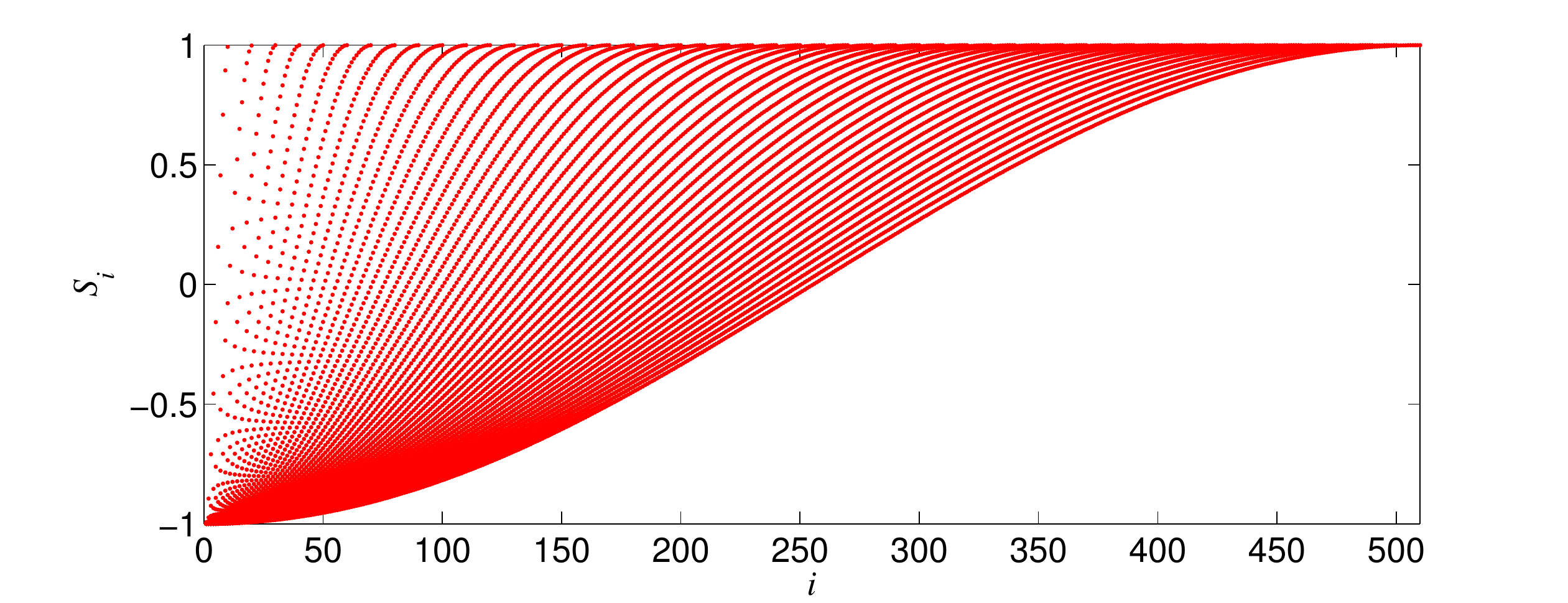}
 \label{fig:HCSQuantizer}
}
\end{center}
\caption{\textbf{(a) Quantized recovery in HCS}. Bernoulli distribution $P(x_i)$ given in Theorem \ref{T:PrRPSign} has estimate $\hat P(x_i)$ (\ref{E:est}) from 1-bit measurements $y=A(x)$. HCS searches the nearest neighbor of $\hat P^(x_i)$ among the $k$ boundaries $P_j$($j=0,\cdots,k-1$) (\ref{E:Bernq}) in the Bernoulli distribution domain. The quantization of $x_i$, i.e., $q_i$ is recovered as the interval between the two boundaries $S_{i-1}$ and $S_i$ corresponding to the nearest neighbor, wherein $S_i$ (\ref{E:HCSq}) is a mapping of $P_{i-1}$ and $P_i$ in signal domain. \textbf{(b) HCS quantizer}. The boundaries $S_i$ in (\ref{E:HCSq}) when $k=10,30,50,\cdots,510$ and $x_{inf}=-1,x_{sup}=1$.}
\end{figure*}

We introduce the HCS quantizer $Q(\cdot)$ by defining a bijective mapping from the boundaries of the Bernoulli distribution domain to the intervals of the signal domain according to Theorem \ref{T:PrRPSign}. Assume the range of a signal $x$ is given by:
\begin{equation}
-1\leq x_{inf}\leq x_i\leq x_{sup}\leq 1,\forall i,\cdots,n.
\end{equation}

By applying the uniform linear quantizer with the quantization interval $\Delta$ to the Bernoulli distribution domain, we get the corresponding boundaries
\begin{equation}\label{E:Bernq}
P_i=\left\{
  \begin{array}{ll}
    P_i^-=\Pr\left(-1\right)=\frac{1}{\pi}\arccos\left(x_{inf}\right)-i\Delta, \\
    P_i^+=\Pr\left(1\right)=1-\Pr\left(-1\right).
  \end{array}
\right.,i=0,\cdots,k-1.
\end{equation}
The interval $\Delta$ is
\begin{equation}
\Delta=\frac{1}{k-1}\left(\frac{1}{\pi}\arccos\left(x_{inf}\right)-\frac{1}{\pi}\arccos\left(x_{sup}\right)\right)=\frac{1}{k-1}\left(P_0^--P_k^-\right).
\end{equation}

We define the k-bit HCS quantizer in the signal domain by computing its $k+1$ boundaries as a mapping from the $k$ boundaries $P_i(i=0,\cdots,k-1)$ to $\mathbb R$ in the Bernoulli domain:
\begin{equation}\label{E:HCSq}
S_i=\left\{
      \begin{array}{ll}
        x_{inf}, &\hbox{$i=0$;} \\
        \cos\left(\frac{\pi}{1+f\left(P_i^-\right)}\right),&\hbox{$i=1,\cdots,k-1$;} \\
        x_{sup}, &\hbox{$i=k$.}
      \end{array}
    \right.
\end{equation}
where
\begin{equation}
\notag f\left(P_i^-\right)=\left(\frac{\left(P_i^-\right)^{\left(P_i^-\right)}\left(1-P_i^-\right)^{\left(1-P_i^-\right)}}{\left(P_i^-+\Delta\right)^{\left(P_i^-+\Delta\right)}\left(1-P_i^--\Delta\right)^{\left(1-P_i^--\Delta\right)}}\right)^{1/\Delta}.
\end{equation}
Although the mapping between the boundaries of HCS quantizer $S_i$ to the boundaries of the quantizer in the Bernoulli distribution domain $P_i$ is bijective, such mapping cannot be explicitly obtained. So it is difficult to derive the corresponding quantizer in the Bernoulli distribution domain from a predefined HCS quantizer. Thus HCS quantizer cannot be fixed as a uniform linear quantizer and has to be computed from a predefined quantizer in the Bernoulli distribution domain. Fortunately, HCS quantizer performs very closely to the uniform linear quantizer, especially when $x_i$ is not very close to $-1$ or $1$. Figure \ref{fig:HCSQuantizer} shows the fact.

Given a signal $x$ and the boundaries defined in (\ref{E:HCSq}), its k-bit HCS quantization $q$ is:
\begin{equation}\label{E:q}
Q(x)=q, q_i=\left\{j:S_{j-1}\leq x_i\leq S_j\right\}.
\end{equation}

\subsection{KL-divergence based nearest neighbor search}

The $k+1$ boundaries of the k-bit HCS quantizer in (\ref{E:HCSq}) define $k$ intervals in $\mathbb R$. Quantized recovery in HCS reconstructs a quantized signal by estimating which interval each dimension of the signal $x$ lies in. The estimation is obtained by a nearest neighbor search in the Bernoulli distribution domain. To be specific, an estimation of $P(x_i)$ given in (\ref{E:PrRPSign}) can be derived from the 1-bit measurements $y$. For each $P(x_i)$, we find its nearest neighbor among the $k$ boundaries $P_j$($j=0,\cdots,k-1$) (\ref{E:Bernq}) in the Bernoulli distribution domain. The interval that $x_i$ lies in is then estimated as the interval of HCS quantizer corresponding to the nearest neighbor. KL-divergence measures the distance between two Bernoulli distributions in the nearest neighbor search.

According to Theorem \ref{T:PrRPSign}, the bijection from $x_i$ to a particular Bernoulli distribution, i.e., $P(x_i)$ given in (\ref{E:PrRPSign}), has an unbiased estimation from the 1-bit measurements $y$
\begin{equation}\label{E:est}
\hat P(x_i)=\left\{
  \begin{array}{ll}
    \hat P(x_i)^-=\hat\Pr\left(s_i=-1\right)=\left|j:\left[y\cdot{\rm sign}\left(\Phi_i'\right)\right]_j=-1\right|/m, \\
    \hat P(x_i)^+=\hat\Pr\left(s_i=1\right)=1-\hat\Pr\left(s_i=-1\right),
  \end{array}
\right.
\end{equation}
where $\Phi_i$ is the $i^{th}$ column of the measurement matrix $\Phi$.

The quantization of $x_i$ can then be recovered by searching the nearest neighbor of $\hat P(x_i)$ among the $k$ boundary Bernoulli distributions $P_j(j=0,\cdots,k-1)$ in (\ref{E:Bernq}). In this paper, the distance between $P_j$ and $\hat P(x_i)$ is measured by the KL-divergence:
\begin{align}\label{E:KLd}
\notag D_{KL}&\left(P_j\|\hat P(x_i)\right)=P_j^-\log\frac{P_j^-}{\hat P(x_i)^-}+P_j^+\log\frac{P_j^+}{\hat P(x_i)^+}, \\
&~~\forall i=1,\cdots,n, \forall j=0,\cdots,k-1.
\end{align}
The interval that $x_i$ lies in among the $k$ intervals defined by the boundaries $S_j(j=0,\cdots,k)$ in (\ref{E:HCSq}) is identified as the one whose corresponding boundary distribution $P_j$ is the nearest neighbor of $\hat P(x_i)$. Therefore, the quantized recovery of $x$, i.e., $q^*$, is given by
\begin{align}\label{E:recover}
\notag R(y)&=q^*, q^*_i=1+\arg\min\limits_{j}D_{KL}\left(P_j\|\hat P(x_i)\right),\\
&\forall i=1,\cdots,n, \forall j=0,\cdots,k-1.
\end{align}
Thus the interval that $x_i$ lies in can be recovered as
\begin{equation}
S_{q^*_i-1}\leq x_i\leq S_{q^*_i}.
\end{equation}
The HCS recovery algorithm is fully summarized in (\ref{E:recover}), which only includes simple computations without iteration and thus can be easily implemented in real systems. According to (\ref{E:recover}), the quantized recovery in HCS requires $nk$ computations of KL-divergence between two Bernoulli distributions. This indicates the high efficiency of HCS (linear recovery time), and the trade-off between resolution ($k$) and time cost ($nk$).

\subsection{Quantized recovery error bound}

We investigate the error bound of the quantized recovery (\ref{E:recover}) by studying the difference between $q_i$ (\ref{E:q}) and $q^*_i$, which are the quantization of $x_i$ and its quantized recovery by HCS, respectively. The difference between $q$ and $q^*$ defines the error $err_H$ in (\ref{E:err}), which is the error caused by HCS reconstruction (\ref{E:recover}):
\begin{equation}
\left|\left(err_H\right)_i\right|=\left\{
                       \begin{array}{ll}
                         S_{q_i}-S_{q^*_i+1}\leq\left(q_i-q^*_i-1\right)\Delta_{max}, & \hbox{$q_i>q^*_i$;} \\
                         0, & \hbox{$q_i=q^*_i$;} \\
                         S_{q^*_i}-S_{q_i+1}\leq\left(q^*_i-q_i-1\right)\Delta_{max}, & \hbox{$q_i<q^*_i$.}
                       \end{array}
                     \right.
\end{equation}
The $\Delta_{max}$ denotes the largest interval between neighboring boundaries of the HCS quantizer, i.e., $\Delta_{max}=\max_{j=1,\cdots,k}\left(S_{j}-S_{j-1}\right)$.

In order to investigate the difference between $q_i$ and $q^*_i$, we study the upper bound for the probability of the event that the true quantization of $x_i$ is $q_i=1+\alpha$, while its recovery by HCS is $q^*_i=1+\beta$($\beta\neq\alpha$). According to the HCS quantizer (\ref{E:q}) and the HCS reconstruction (\ref{E:recover}), this probability is
\begin{equation}\label{E:failureP}
\Pr\left(\beta=\arg\min\limits_{j}D_{KL}\left(P_j\|\hat P(x_i)\right)\mid S_{\alpha}\leq x_i\leq S_{\alpha+1}\right).
\end{equation}
In order to study the conditional probability in (\ref{E:failureP}), we first consider an equivalent event of $\beta=\arg\min\limits_{j}D_{KL}\left(P_j\|\hat P(x_i)\right)$, shown in the following Lemma \ref{L:euqal}.

\begin{lemma}\label{L:euqal}
{\rm(\textbf{Equivalence})} The event that the nearest neighbor of $\hat P(x_i)$ among $P_j(j=0,\cdots,k-1)$ is $P_\beta$ equals to the event that $\hat P(x_i)$ is closer to $P_\beta$ than both $P_{\beta-1}$ and $P_{\beta+1}$, where the distance between $P_j$ and $\hat P(x_i)$ is measured by KL divergence (\ref{E:KLd}), i.e.,
\begin{align}\label{E:euqal}
\notag &\beta=\arg\min\limits_{j}D_{KL}\left(P_j\|\hat P(x_i)\right)\Longleftrightarrow\\
&\left\{
   \begin{array}{ll}
     D_{KL}\left(P_{\beta-1}\|\hat P(x_i)\right)-D_{KL}\left(P_\beta\|\hat P(x_i)\right)>0, \\
     D_{KL}\left(P_{\beta+1}\|\hat P(x_i)\right)-D_{KL}\left(P_\beta\|\hat P(x_i)\right)>0.
   \end{array}
 \right.
\end{align}
\end{lemma}
\begin{proof}
It is direct to have the following equivalence:
\begin{equation}
\beta=\arg\min\limits_{j}D_{KL}\left(P_j\|\hat P(x_i)\right)\Longleftrightarrow D_{KL}\left(P_{j:j\neq\beta}\|\hat P(x_i)\right)-D_{KL}\left(P_\beta\|\hat P(x_i)\right)>0.
\end{equation}
Thus ``$\Longrightarrow$'' in (\ref{E:euqal}) is true. In order to prove ``$\Longleftarrow$'' in (\ref{E:euqal}), for arbitrary $j\in\{0,\cdots,k-1\}$ and fixed $x_i$, we study the monotonicity of $D_{KL}\left(P_{j}\|\hat P(x_i)\right)$ as a function of $P_j^-$:
\begin{equation}
\frac{\partial D_{KL}\left(P_{j}\|\hat P(x_i)\right)}{\partial P_j^-}=\log\left(\frac{P_j^-}{\hat P(x_i)^-}\cdot\frac{1-\hat P(x_i)^-}{1-P_j^-}\right).
\end{equation}
Therefore, it holds that
\begin{equation}
\frac{\partial D_{KL}\left(P_{j}\|\hat P(x_i)\right)}{\partial P_j^-}\left\{
                                                                       \begin{array}{ll}
                                                                         >0, & \hbox{$P_j^->\hat P(x_i)^-$;} \\
                                                                         <0, & \hbox{$P_j^-<\hat P(x_i)^-$.}
                                                                       \end{array}
                                                                     \right.
\end{equation}
According to the definition of $P_j$ in (\ref{E:Bernq}) and the right hand side of ``$\Longleftrightarrow$'' in (\ref{E:euqal}), we have $P_{j:j=0,\cdots,\beta-1}^->\hat P(x_i)^-$ that indicate
\begin{equation}
D_{KL}\left(P_{j:j=0,\cdots,\beta-2}\|\hat P(x_i)\right)>D_{KL}\left(P_{\beta-1}\|\hat P(x_i)\right)>D_{KL}\left(P_{\beta}\|\hat P(x_i)\right),
\end{equation}
and $P_{j:j=\beta+1,\cdots,k-1}^-<\hat P(x_i)^-$ that indicate
\begin{equation}
D_{KL}\left(P_{j:j=\beta+1,\cdots,k-1}\|\hat P(x_i)\right)>D_{KL}\left(P_{\beta+1}\|\hat P(x_i)\right)>D_{KL}\left(P_{\beta}\|\hat P(x_i)\right).
\end{equation}
Therefore, we can derive the left hand side of ``$\Longleftrightarrow$'' from its right hand side in (\ref{E:euqal}). This completes the proof.
\end{proof}

By using the equivalence in Lemma \ref{L:euqal}, the conditional probability given in (\ref{E:failureP}) can be upper bounded by two other conditional probabilities, whose conditions are the two cases of the condition in (\ref{E:failureP}).

\begin{corollary}\label{C:2P}
{\rm(\textbf{Upper bounds in two cases})} The conditional probability given in (\ref{E:failureP}) can be upper bounded by
\begin{align}\label{E:2P}
\notag&\Pr\left(\beta=\arg\min\limits_{j}D_{KL}\left(P_j\|\hat P(x_i)\right)\mid S_{\alpha}\leq x_i\leq S_{\alpha+1}\right)\\
\leq&\left\{
    \begin{array}{ll}
      \Pr\left(D_{KL}\left(P_{\beta-1}\|\hat P(x_i)\right)-D_{KL}\left(P_\beta\|\hat P(x_i)\right)>0\mid S_{\alpha}\leq x_i\leq S_{\alpha+1}\leq S_\beta\right), \\
      \Pr\left(D_{KL}\left(P_{\beta+1}\|\hat P(x_i)\right)-D_{KL}\left(P_\beta\|\hat P(x_i)\right)>0\mid S_{\beta+1}\leq S_{\alpha}\leq x_i\leq S_{\alpha+1}\right).
    \end{array}
  \right.
\end{align}
\end{corollary}
\begin{proof}
By using Lemma \ref{L:euqal}, we discuss the the conditional probability in (\ref{E:failureP}) by considering the two cases of the conditional event $S_{\alpha}\leq x_i\leq S_{\alpha+1}$.

Case 1) When $S_{\alpha+1}\leq S_\beta$, we have
\begin{align}
\notag&\Pr\left(\beta=\arg\min\limits_{j}D_{KL}\left(P_j\|\hat P(x_i)\right)\mid S_{\alpha}\leq x_i\leq S_{\alpha+1}\right)\\
\notag =&\Pr\left(D_{KL}\left(P_{\beta-1}\|\hat P(x_i)\right)-D_{KL}\left(P_\beta\|\hat P(x_i)\right)>0\mid S_{\alpha}\leq x_i\leq S_{\alpha+1}\leq S_\beta\right)\cdot \\
\notag&\Pr\left(D_{KL}\left(P_{\beta+1}\|\hat P(x_i)\right)-D_{KL}\left(P_\beta\|\hat P(x_i)\right)>0\mid S_{\alpha}\leq x_i\leq S_{\alpha+1}\leq S_\beta\right)\\
\leq&\Pr\left(D_{KL}\left(P_{\beta-1}\|\hat P(x_i)\right)-D_{KL}\left(P_\beta\|\hat P(x_i)\right)>0\mid S_{\alpha}\leq x_i\leq S_{\alpha+1}\leq S_\beta\right).
\end{align}

Case 2) When $S_{\beta+1}\leq S_{\alpha}$, we have
\begin{align}
\notag&\Pr\left(\beta=\arg\min\limits_{j}D_{KL}\left(P_j\|\hat P(x_i)\right)\mid S_{\alpha}\leq x_i\leq S_{\alpha+1}\right)\\
\notag=&\Pr\left(D_{KL}\left(P_{\beta-1}\|\hat P(x_i)\right)-D_{KL}\left(P_\beta\|\hat P(x_i)\right)>0\mid S_{\beta+1}\leq S_{\alpha}\leq x_i\leq S_{\alpha+1}\right)\cdot\\
\notag&\Pr\left(D_{KL}\left(P_{\beta+1}\|\hat P(x_i)\right)-D_{KL}\left(P_\beta\|\hat P(x_i)\right)>0\mid S_{\beta+1}\leq S_{\alpha}\leq x_i\leq S_{\alpha+1}\right)\\
\leq&\Pr\left(D_{KL}\left(P_{\beta+1}\|\hat P(x_i)\right)-D_{KL}\left(P_\beta\|\hat P(x_i)\right)>0\mid S_{\beta+1}\leq S_{\alpha}\leq x_i\leq S_{\alpha+1}\right).
\end{align}

This completes the proof.
\end{proof}

Hence we can bound the conditional probability in (\ref{E:failureP}) by exploring the upper bounds of the two conditional probabilities in Corollary \ref{C:2P}.
\begin{proposition}\label{P:2bounds}
{\rm(\textbf{Two probabilistic bounds})} The two conditional probabilities in (\ref{E:2P}) are upper bounded by
\begin{align}
\notag &\Pr\left(D_{KL}\left(P_{\beta-1}\|\hat P(x_i)\right)-D_{KL}\left(P_\beta\|\hat P(x_i)\right)>0\mid S_{\alpha}\leq x_i\leq S_{\alpha+1}\leq S_\beta\right)\leq\\
&\frac{1}{2}\exp\left(-2m\cdot\left(\frac{1}{\pi}\arccos\left(x_i\right)-\frac{1}{f\left(P_\beta^-\right)+1}\right)^2\right),\\
&\notag \Pr\left(D_{KL}\left(P_{\beta+1}\|\hat P(x_i)\right)-D_{KL}\left(P_\beta\|\hat P(x_i)\right)>0\mid S_{\beta+1}\leq S_{\alpha}\leq x_i\leq S_{\alpha+1}\right)\leq\\
&\frac{1}{2}\exp\left(-2m\cdot\left(\frac{1}{f\left(P_{\beta+1}^-\right)+1}-\frac{1}{\pi}\arccos\left(x_i\right)\right)^2\right),
\end{align}
where $f$ is defined as
\begin{equation}
f\left(P_j^-\right)=\left(\frac{\left(P_j^-\right)^{\left(P_j^-\right)}\left(1-P_j^-\right)^{\left(1-P_j^-\right)}}{\left(P_j^-+\Delta\right)^{\left(P_j^-+\Delta\right)}\left(1-P_j^--\Delta\right)^{\left(1-P_j^--\Delta\right)}}\right)^{1/\Delta}.
\end{equation}
\end{proposition}
\begin{proof}
For proving (27), according to (\ref{E:KLd}) and the definition of $\hat P(x_i)^-$ in (\ref{E:est}), we have the following equivalences:
\begin{align}
\notag &D_{KL}\left(P_{\beta-1}\|\hat P(x_i)\right)-D_{KL}\left(P_\beta\|\hat P(x_i)\right)\\
\notag =&\Delta\log\frac{1-\hat P(x_i)^-}{\hat P(x_i)^-}-\log\frac{\left(P_\beta^-\right)^{\left(P_\beta^-\right)}\left(1-P_\beta^-\right)^{\left(1-P_\beta^-\right)}}{\left(P_\beta^-+\Delta\right)^{\left(P_\beta^-+\Delta\right)}\left(1-P_\beta^--\Delta\right)^{\left(1-P_\beta^--\Delta\right)}}>0\Longleftrightarrow\\
\notag &\hat P(x_i)^-<\frac{1}{f\left(P_\beta^-\right)+1},f\left(P_j^-\right)=\left(\frac{\left(P_j^-\right)^{\left(P_j^-\right)}\left(1-P_j^-\right)^{\left(1-P_j^-\right)}}{\left(P_j^-+\Delta\right)^{\left(P_j^-+\Delta\right)}\left(1-P_j^--\Delta\right)^{\left(1-P_j^--\Delta\right)}}\right)^{1/\Delta}\Longleftrightarrow\\
&\left|j:\left[y\cdot{\rm sign}\left(\Phi_i'\right)\right]_j=-1\right|\in\left[0,\underline{\frac{m}{f\left(P_\beta^-\right)+1}}\right],
\end{align}
where $\underline{x}$ denotes the largest integer smaller than $x$.

Since $\left|j:\left[y\cdot{\rm sign}\left(\Phi_i'\right)\right]_j=-1\right|$ refers to the event that in a sequence of $m$ independent Bernoulli trials defined in (\ref{E:PrRPSign}), there are $j$ trials return $s_i=-1$, we can conclude the distribution of $j$ follows the binomial distribution
\begin{equation}\label{E:binomial}
\Pr\left(\left|j:\left[y\cdot{\rm sign}\left(\Phi_i'\right)\right]_j=-1\right|\right)={m\choose j}\left(\frac{1}{\pi}\arccos\left(x_i\right)\right)^j\left(1-\frac{1}{\pi}\arccos\left(x_i\right)\right)^{m-j}.
\end{equation}
According to the equivalence shown in (30), the probability in (27) can then be computed as
\begin{equation}\label{E:Pf1}
\Pr\left(j\in\left[0,\underline{\frac{m}{f\left(P_\beta^-\right)+1}}\right]\right)=\sum\limits_{j=0}^{\underline{\frac{m}{f\left(P_\beta^-\right)+1}}}{m\choose j}\left(\frac{1}{\pi}\arccos\left(x_i\right)\right)^j\left(1-\frac{1}{\pi}\arccos\left(x_i\right)\right)^{m-j}.
\end{equation}
Since
\begin{equation}\label{E:fmonotonicity}
\frac{\partial f\left(P_j^-\right)}{\partial P_j^-}=\log\left(\frac{P_j^-}{P_j^-+\Delta}\cdot\frac{1-P_j^--\Delta}{1-P_j^-}\right)<0,
\end{equation}
we have
\begin{equation}
S_{\alpha+1}\leq S_\beta\Longleftrightarrow{\underline{\frac{m}{f\left(P_\beta^-\right)+1}}}\leq\frac{m}{f\left(P_{\alpha+1}^-\right)+1}=m\cdot\frac{1}{\pi}\arccos\left(S_{\alpha+1}\right).
\end{equation}
Hence the condition of Hoeffding's inequality for probability (\ref{E:Pf1}) holds:
\begin{equation}
{\underline{\frac{m}{f\left(P_\beta^-\right)+1}}}\leq m\cdot\frac{1}{\pi}\arccos\left(S_{\alpha+1}\right)\leq m\cdot\frac{1}{\pi}\arccos\left(x_i\right).
\end{equation}
By applying Hoeffding's inequality to probability (\ref{E:Pf1}), we have
\begin{equation}
\Pr\left(j\in\left[0,\underline{\frac{m}{f\left(P_\beta^-\right)+1}}\right]\right)\leq\frac{1}{2}\exp\left(-2m\cdot\left(\frac{1}{\pi}\arccos\left(x_i\right)-\frac{1}{f\left(P_\beta^-\right)+1}\right)^2\right).
\end{equation}
Due to the equivalence proved in (30), we obtain (27). This completes the proof of (27).

To prove (28), similarly, according to (\ref{E:KLd}) and the definition of $\hat P(x_i)^-$ in (\ref{E:est}), we have the following equivalences:
\begin{align}
\notag &D_{KL}\left(P_{\beta+1}\|\hat P(x_i)\right)-D_{KL}\left(P_\beta\|\hat P(x_i)\right)\\
\notag =&\log\frac{\left(P_{\beta+1}^-\right)^{\left(P_{\beta+1}^-\right)}\left(1-P_{\beta+1}^-\right)^{\left(1-P_{\beta+1}^-\right)}}{\left(P_{\beta+1}^-+\Delta\right)^{\left(P_{\beta+1}^-+\Delta\right)}\left(1-P_{\beta+1}^--\Delta\right)^{\left(1-P_{\beta+1}^--\Delta\right)}}-\Delta\log\frac{1-\hat P(x_i)^-}{\hat P(x_i)^-}>0\Longleftrightarrow\\
&\hat P(x_i)^->\frac{1}{f^-\left(P_{\beta+1}^-\right)+1}\Longleftrightarrow\left|j:\left[y\cdot{\rm sign}\left(\Phi_i'\right)\right]_j=-1\right|\in\left[\overline{\frac{m}{f^-\left(P_{\beta+1}^-\right)+1}},m\right],
\end{align}
where $\overline{x}$ denotes the smallest integer larger than $x$.

According to the equivalence shown in (37) and the binomial distribution given in (\ref{E:binomial}), the probability in (28) can be computed as
\begin{equation}\label{E:Pf2}
\Pr\left(j\in\left[\overline{\frac{m}{f^-\left(P_{\beta+1}^-\right)+1}},m\right]\right)=\sum\limits_{j=0}^{m-\overline{\frac{m}{f\left(P_{\beta+1}^-\right)+1}}}{m\choose j}\left(1-\frac{1}{\pi}\arccos\left(x_i\right)\right)^j\left(\frac{1}{\pi}\arccos\left(x_i\right)\right)^{m-j}.
\end{equation}
The monotonicity of $f\left(P_j^-\right)$ in (\ref{E:fmonotonicity}) yields
\begin{equation}
S_{\beta+1}\leq S_\alpha\Longleftrightarrow m-{\overline{\frac{m}{f\left(P_{\beta+1}^-\right)+1}}}\leq m-\frac{m}{f\left(P_\alpha^-\right)+1}=m\left(1-\frac{1}{\pi}\arccos\left(S_{\alpha}\right)\right).
\end{equation}
Hence the condition of Hoeffding's inequality for probability (\ref{E:Pf2}) holds:
\begin{equation}
m-{\overline{\frac{m}{f\left(P_{\beta+1}^-\right)+1}}}\leq m\left(1-\frac{1}{\pi}\arccos\left(S_{\alpha}\right)\right)\leq m\left(1-\frac{1}{\pi}\arccos\left(x_i\right)\right).
\end{equation}
By applying Hoeffding's inequality to probability (\ref{E:Pf2}), we have
\begin{equation}
\Pr\left(j\in\left[\overline{\frac{m}{f^-\left(P_{\beta+1}^-\right)+1}},m\right]\right)\leq\frac{1}{2}\exp\left(-2m\cdot\left(\frac{1}{f\left(P_{\beta+1}^-\right)+1}-\frac{1}{\pi}\arccos\left(x_i\right)\right)^2\right).
\end{equation}
Due to the equivalence proved in (37), we obtain (28). This completes the proof of (28).
\end{proof}

By using Lemma \ref{L:euqal}, Corollary \ref{C:2P} and Proposition \ref{P:2bounds}, we have the following Theorem about the upper bound of the probability in (\ref{E:failureP}).

\begin{theorem}\label{T:HCSbound}
{\rm(\textbf{Quantized recovery bound})} Given HCS quantizer $Q(\cdot)$ in (\ref{E:q}) and HCS reconstruction $R(\cdot)$ (\ref{E:recover}), the probability of the event that the true quantization of $x_i$ is $q_i=1+\alpha$ while its recovery by HCS is $q^*_i=1+\beta$($q_i\neq q^*_i$) is upper bounded by
\begin{align}\label{E:failurePbound}
\notag &\Pr\left(\left[Q(x)\right]_i=q_i\mid \left[R(y)\right]_i=q^*_i\right)=\Pr\left(\beta=\arg\min\limits_{j}D_{KL}\left(P_j\|\hat P(x_i)\right)\mid S_{\alpha}\leq x_i\leq S_{\alpha+1}\right)\\
&\leq\left\{
       \begin{array}{ll}
         \frac{1}{2}\exp\left(-2m\cdot\left(\frac{1}{f\left(P_{{q^*_i}+1}^-\right)+1}-\frac{1}{\pi}\arccos\left(x_i\right)\right)^2\right), & \hbox{$q_i>q^*_i$;}\\
         \frac{1}{2}\exp\left(-2m\cdot\left(\frac{1}{\pi}\arccos\left(x_i\right)-\frac{1}{f\left(P_{q^*_i}^-\right)+1}\right)^2\right), & \hbox{$q_i< q^*_i$.}
       \end{array}
     \right.
\end{align}
\end{theorem}

The minimum amount of 1-bit measurements that ensures the successful quantized recovery in HCS is then directly obtained from Theorem \ref{T:HCSbound}.

\begin{corollary}\label{C:AmountM}
{\rm(\textbf{Amount of measurements})} HCS successfully reconstructs $x_i$ with probability exceeding $1-\eta$ ($0\leq\eta\leq 1$) if the number of measurements
\begin{equation}\label{E:measurebound}
m\geq\frac{1}{2\delta_i}\log\frac{1}{2\eta},
\end{equation}
where
\begin{equation}\label{E:delta}
\delta_i=\min\left[\left(\frac{1}{f\left(P_{q_i}^-\right)+1}-\frac{1}{\pi}\arccos\left(x_i\right)\right)^2,\left(\frac{1}{\pi}\arccos\left(x_i\right)-\frac{1}{f\left(P_{q_i+1}^-\right)+1}\right)^2\right].
\end{equation}
Moreover, HCS successfully reconstruct the signal $x$ with probability exceeding $1-\eta$  if the number of measurements
\begin{equation}\label{E:overallMbound}
m\geq\frac{1}{2\min_i\delta_i}\log\frac{n}{2\eta}.
\end{equation}
\end{corollary}

\textbf{Remark:} Corollary \ref{C:AmountM} states that the quantization of an $n$-dimensional signal $x$ on the unit sphere can be successfully recovered by HCS from $m=\mathcal O(\log n)$ with high probability. Compared with CS and QCS, the amount of measurements required by HCS is substantially reduced and irrelevant to the sparsity of the signal. Thus HCS provides a simpler and more economical sampling scheme that does not rely on sparse or compressible assumption to the signal.

A new issue in quantized recovery is the influence of quantization bits $k$ to the recovery accuracy. According to the definition of $\delta_i$ in (\ref{E:delta}), both the upper bound for the probability of reconstruction failure in (\ref{E:failurePbound}) and the least number of measurements ensuring reconstruction success in (\ref{E:measurebound}) will be reduced if $\left|q_i-q^*_i\right|$ increases. This indicates two facts: 1) the interval $x_i$ lies in is easier to be mistakenly recovered as its nearest intervals; and 2) when we increase the number of bits $k$ in quantized recovery, $x_i$ will become closer to the boundaries $S_{q-1}$ and $S_q$, which leads to the decreasing of $\min_i\delta_i$ in (\ref{E:overallMbound}). In this case, the number of measurements $m$ has to be increased in order to ensure a successful recovery. In summary, recovering finer quantization in HCS requires an augment in the number of measurements. In another word, HCS performs a trade-off between sampling rate and resolution.

\section{Dequantizer}

If required, we can dequantize the quantized recovery of the signal by assigning to $x^*_i$ the midpoint of the interval that $x_i$ lies in, i.e., $ x^*_i=\frac{1}{2}\left(S_{q^*_i-1}+S_{q^*_i}\right)$. Although this dequantizer is simple and efficient, it is not accurate.

Fortunately, existing 1-bit CS provides accurate tools for dequantization on the HCS quantized recovery result compared with the midpoint reconstruction, though they introduce extra computational costs to trade-off the efficiency against dequantization accuracy. That is because most 1-bit CS recovery algorithms invoke time consuming optimization with $\ell_p$($0\leq p\leq 2$) penalty or constraint. However, the quantized recovery of HCS provides a box constraint to the subsequent 1-bit CS optimization, and thus significantly reduces the time costs by shrinking the search space for $x^*$.

In particular, we obtain the following box constraint to the signal $x$ from the HCS quantized recovery $q^*$:
\begin{equation}\label{E:box}
\Omega=\left\{x:S_{q^*_i-1}\leq x_i\leq S_{q^*_i}\right\}.
\end{equation}

Since $\Omega$ in (\ref{E:box}) is convex, the projection to it is direct
\begin{equation}\label{E:proj}
\mathcal P_{\Omega}\left(x\right)=z, z_i={\rm median}\left\{S_{q^*_i-1},x_i,S_{q^*_i}\right\},
\end{equation}

The dequantization can then be obtained by adding a projection step (\ref{E:proj}) at the end of each iteration round of 1-bit CS recovery algorithms. Note the 1-bit CS algorithms with this modification have a substantially smaller searching space, so they will converge quickly. Note $x^*$ has to be normalized to $x^*:=x^*/\|x^*\|_2$ at the end of the dequantization, because $x$ is assumed to be on the unit $\ell_2$ sphere.

\subsection{HCS+dequantizer error bound}

We analyze the error of ``HCS+dequantizer'' recovery based on the fact that both the error $err_H$ caused by HCS reconstruction and the error $err_D$ caused by a dequantizer can be upper bounded. In the worst case, the upper bound of the dequantization error $err_D$ is
\begin{equation}
\left|\left(err_D\right)_i\right|\leq S_{q^*_i}-S_{q^*_i-1}, \forall i=1,\cdots,n.
\end{equation}

Based on the consistency in Lemma \ref{L:Consis}, the definition of B$\epsilon$SE \cite{Robust1bitCS} and Lemma of amount of measurement \cite{Robust1bitCS}, we derive the upper bound of ``HCS+dequantizer'' recovery error when the signal is $K$-sparse in the following Theorem \ref{T:Dbound}.

\begin{definition}\label{D:BeSE}
{\rm(\textbf{Binary $\epsilon$-Stable Embedding})} Let $\epsilon\in(0,1)$. A mapping $A:\mathbb R^n\rightarrow\{-1,1\}^m$ is a binary $\epsilon$-stable embedding (B$\epsilon$SE) of order $K$ for sparse signals if
\begin{equation}\label{E:BeSE}
D_S(x,x^*)-\epsilon\leq D_H\left(A(x),A(x^*)\right)\leq D_S(x,x^*)+\epsilon, D_S(x,x^*)=\frac{1}{\pi}\arccos\langle x,x^*\rangle.
\end{equation}
for all $K$-sparse signals $x,x^*$ on the unit $\ell_2$ sphere.
\end{definition}

\begin{lemma}\label{L:AmountM}
{\rm(\textbf{Amount of measurements})} Let $\Phi$ be the measurement matrix defined in Theorem \ref{T:PrRPSign}, and let the 1-bit measurement operator $A(\cdot)$ be defined as in (\ref{E:1bitm}). Fix $0\leq\mu\leq 1$ and $\epsilon>0$. If the number of measurements is
\begin{equation}\label{E:AmountM}
m\geq\frac{4}{\epsilon^2}\left(K\log n+2K\log\frac{50}{\epsilon}+\log\frac{2}{\mu}\right),
\end{equation}
then with probability exceeding $1-\mu$, the mapping defined by $A(\cdot)$ is a B$\epsilon$SE of order $K$ for sparse signals.
\end{lemma}

\begin{theorem}\label{T:Dbound}
{\rm(\textbf{HCS+dequantizer error bound})} If $x^*=D(q^*)=D(R(y))=D(R(A(x)))$ is the ``HCS+dequantizer'' recovery of $K$-sparse signal $x$, where $y$ includes $m$ measurements whose amount satisfies (\ref{E:AmountM}), then
\begin{equation}
D_S\left(x,x^*\right)\leq D_H\left(A(x),A(x^*)\right)+\epsilon\leq\frac{1}{2}\frac{\sigma}{\|x\|_2}+\gamma+\epsilon,
\end{equation}
where $\sigma$, $\gamma$ are defined in Lemma \ref{L:Consis}.
\end{theorem}

\begin{figure}[htb]
\begin{center}
\subfigure{
 \includegraphics[width=1\linewidth]{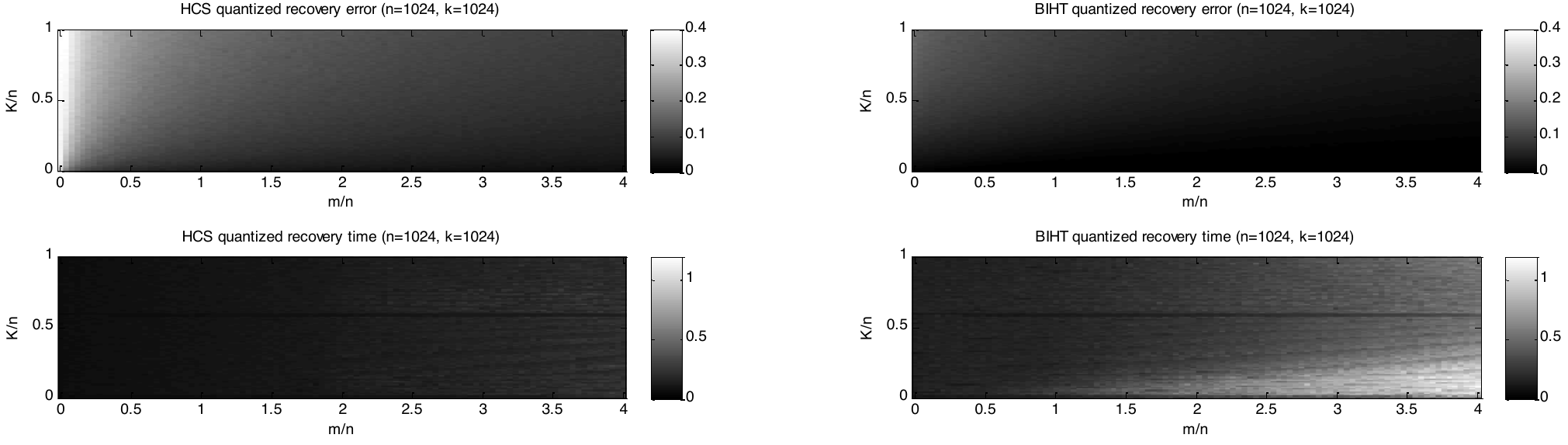}
}
\subfigure{
 \includegraphics[width=1\linewidth]{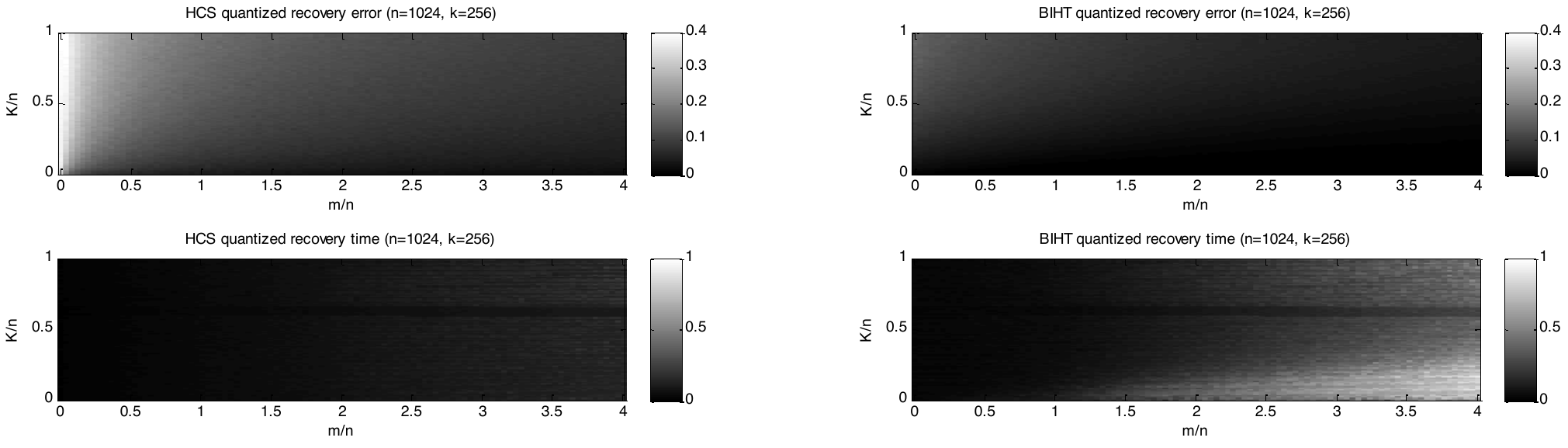}
}
\subfigure{
 \includegraphics[width=1\linewidth]{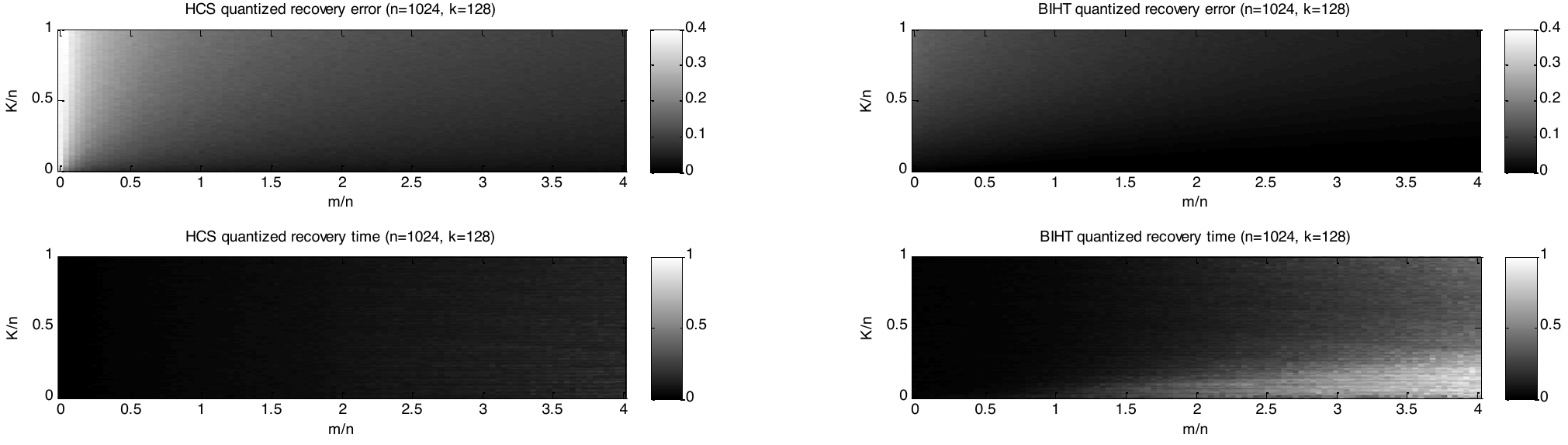}
}
\end{center}
 \caption{Phase plots of HCS and ``1-bit CS+HCS quantizer'' in the noiseless case.}
\label{fig:phase1}
\end{figure}

\section{Empirical study}

This section compares HCS and compare it with BIHT \cite{Robust1bitCS} for 1-bit CS on 3 groups of numerical experiments. We use average quantized recovery error $\sum_{i=1}^n\left|q_i-q^*_i\right|/nk$ to measure $err_H$ shown in Section 3.3. In each trial, we draw a normalized Gaussian random matrix $\Phi\in\mathbb R^{m\times n}$ given in Theorem \ref{T:PrRPSign} and a signal of length $n$ and cardinality $K$, whose $K$ nonzero entries drawn uniformly at random on the unit $\ell_2$ sphere.

\subsection{Phase transition in the noiseless case}

\begin{figure}[htb]
\begin{center}
\subfigure{
 \includegraphics[width=1\linewidth]{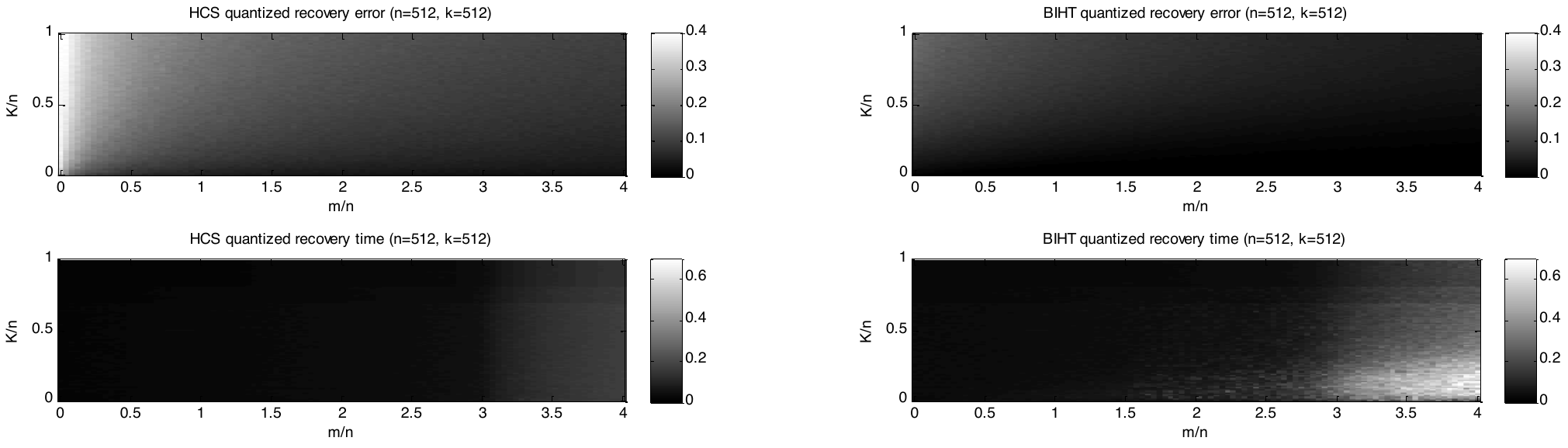}
}
\subfigure{
 \includegraphics[width=1\linewidth]{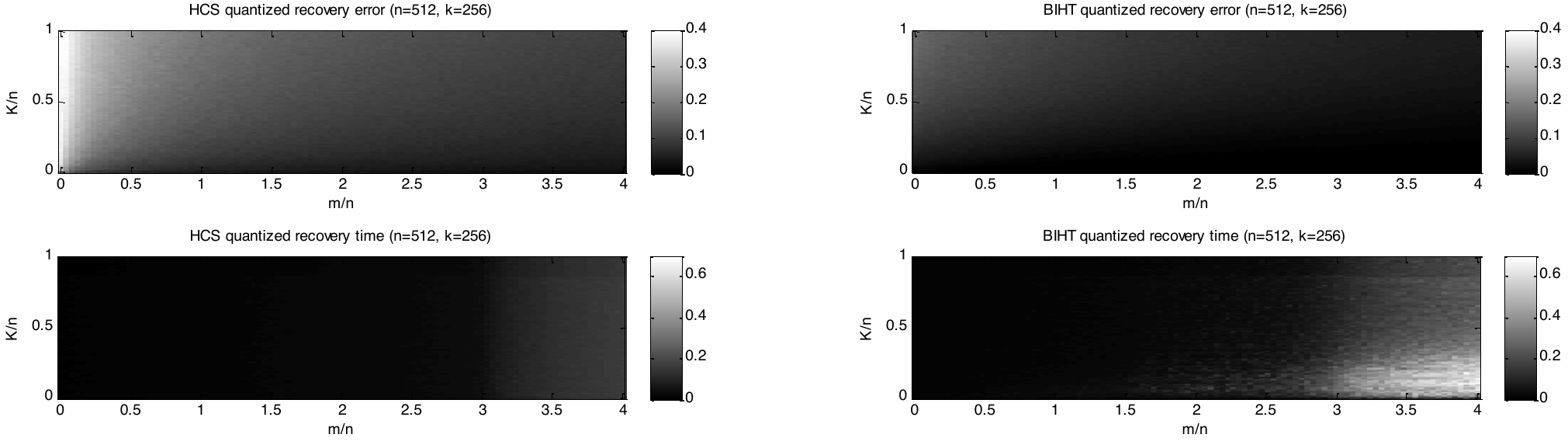}
}
\subfigure{
 \includegraphics[width=1\linewidth]{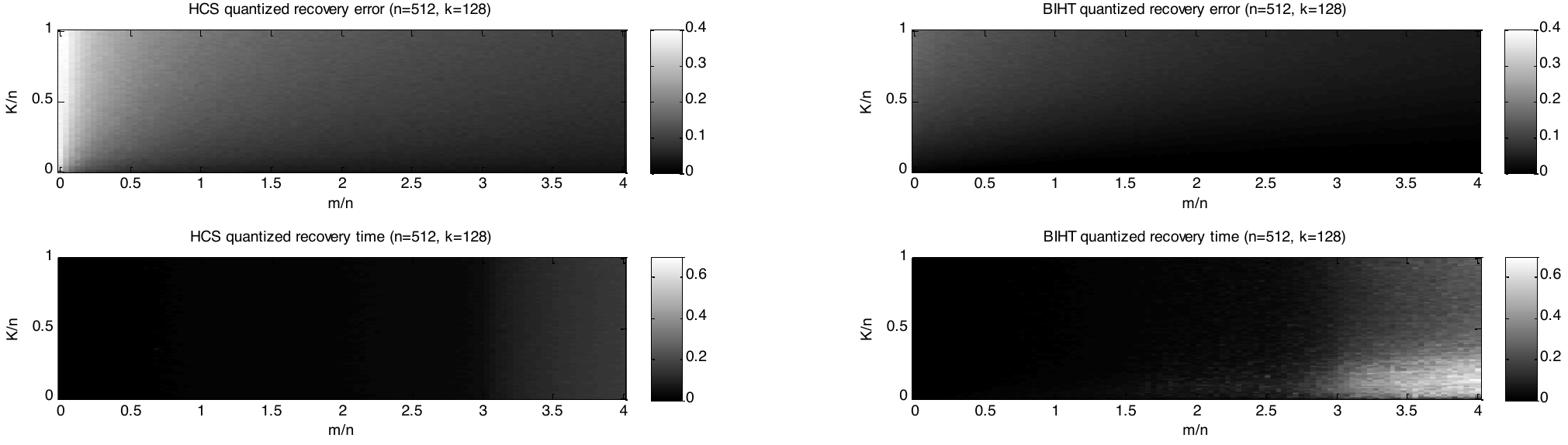}
}
\end{center}
 \caption{Phase plots of HCS and ``1-bit CS+HCS quantizer'' in the noiseless case.}
\label{fig:phase2}
\end{figure}

We first study the phase transition properties of HCS and 1-bit CS on quantized recovery error and on recovery time in the noiseless case. We conduct HCS and ``BIHT+HCS quantizer'' for $10^5$ trials. In particular, given fixed $n$ and $k$, we uniformly choose $100$ different $K/n$ values between $0$ and $1$, and $100$ different $m/n$ values between $0$ and $4$. For each $\{K/n,m/n\}$ pair, we conduct $10$ trials, i.e., HCS recovery and ``1-bit CS+HCS quantizer'' of $10$ $n$-dimensional signals with cardinality $K$ from their $m$ 1-bit measurements. The average quantized recovery errors and average time costs of the two methods on overall $10^4$ $\{K/n,m/n\}$ pairs for different $n$ and $k$ are shown in Figure \ref{fig:phase1} and Figure \ref{fig:phase2}.

In Figure \ref{fig:phase1} and Figure \ref{fig:phase2}, the phase plots of quantized recovery error show the quantized recovery of HCS is accurate if the 1-bit measurements are sufficient. Compared to ``1-bit CS+HCS quantizer'', HCS needs slightly more measurements to reach the same recovery precision. That is because 1-bit CS recovers the exact signal, while HCS recovers its quantization. Another reason is HCS quantizer performs different from uniform linear quantizer when $x_i$ approaching $-1$ or $1$ for the normalized signal $x$, which exactly corresponds to the left margin area of the phase plot. However, the phase plots of quantized recovery time shows that HCS costs substantially less time than ``1-bit CS+HCS quantizer''. Thus HCS can significantly improve the efficiency of practical digital systems and eliminate the hardware cost for additional quantization.

\subsection{Phase transition in the noisy case}

\begin{figure}[htb]
\begin{center}
\subfigure{
 \includegraphics[width=1\linewidth]{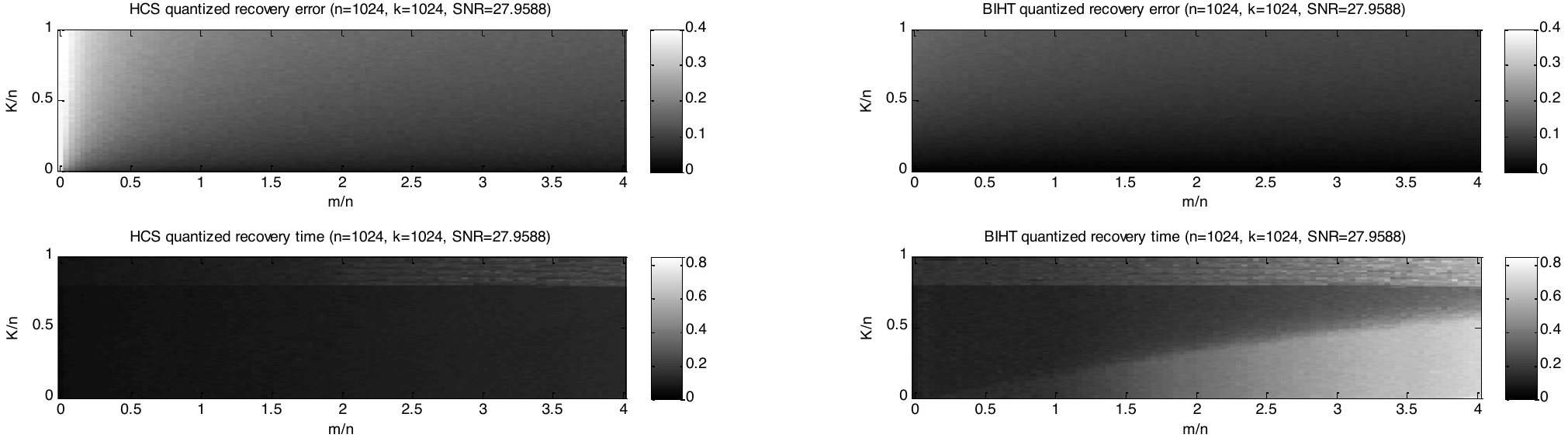}
}
\subfigure{
 \includegraphics[width=1\linewidth]{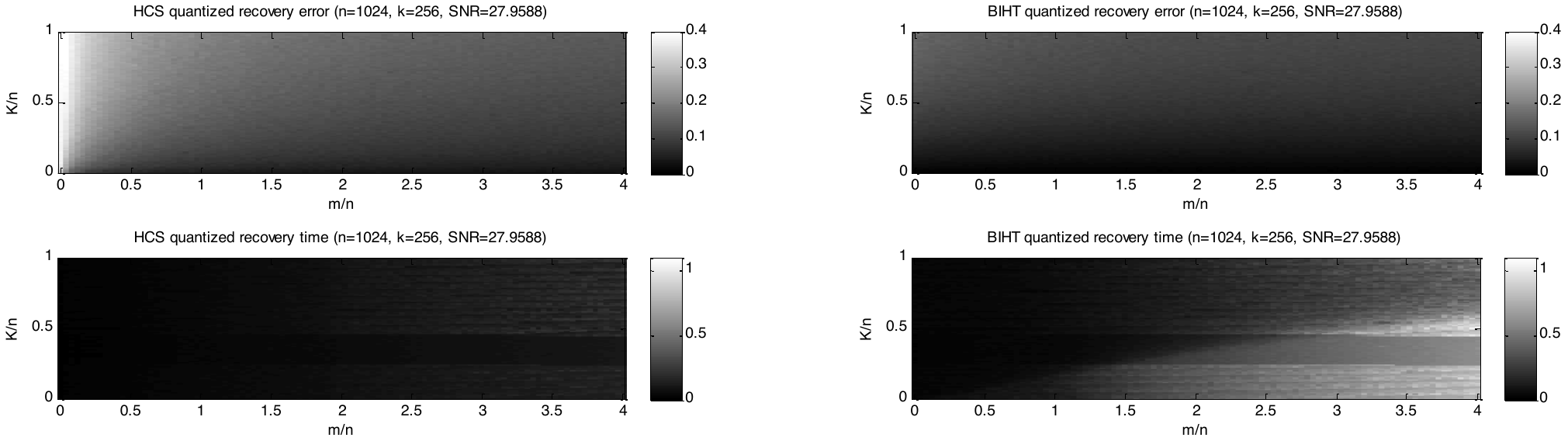}
}
\subfigure{
 \includegraphics[width=1\linewidth]{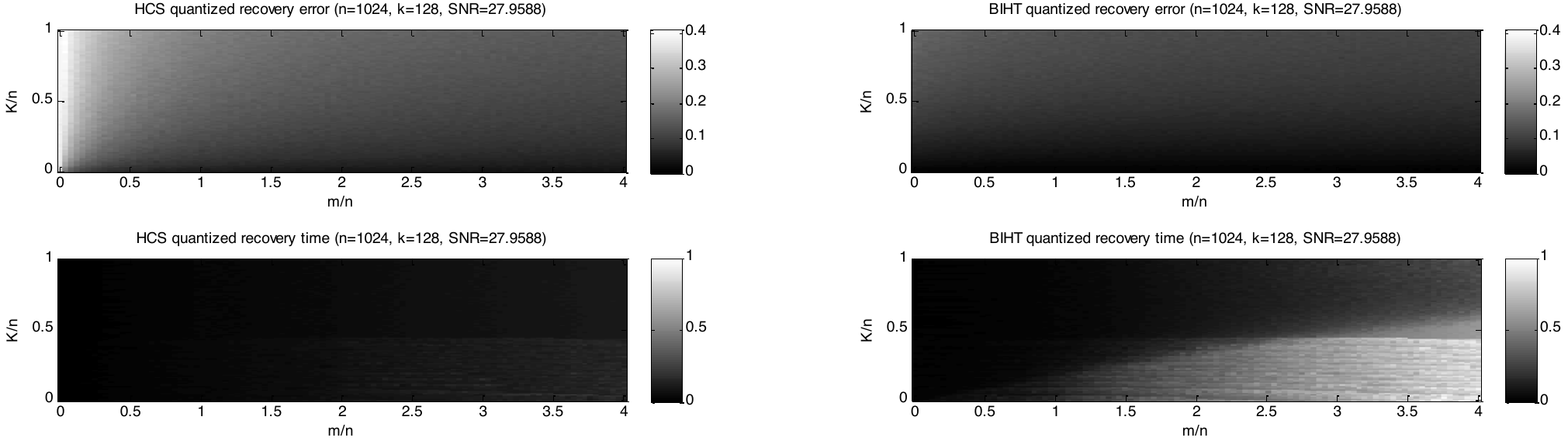}
}
\end{center}
 \caption{Phase plots of HCS and ``1-bit CS+HCS quantizer'' in the noisy case.}
\label{fig:nphase1}
\end{figure}

\begin{figure}[htb]
\begin{center}
\subfigure{
 \includegraphics[width=1\linewidth]{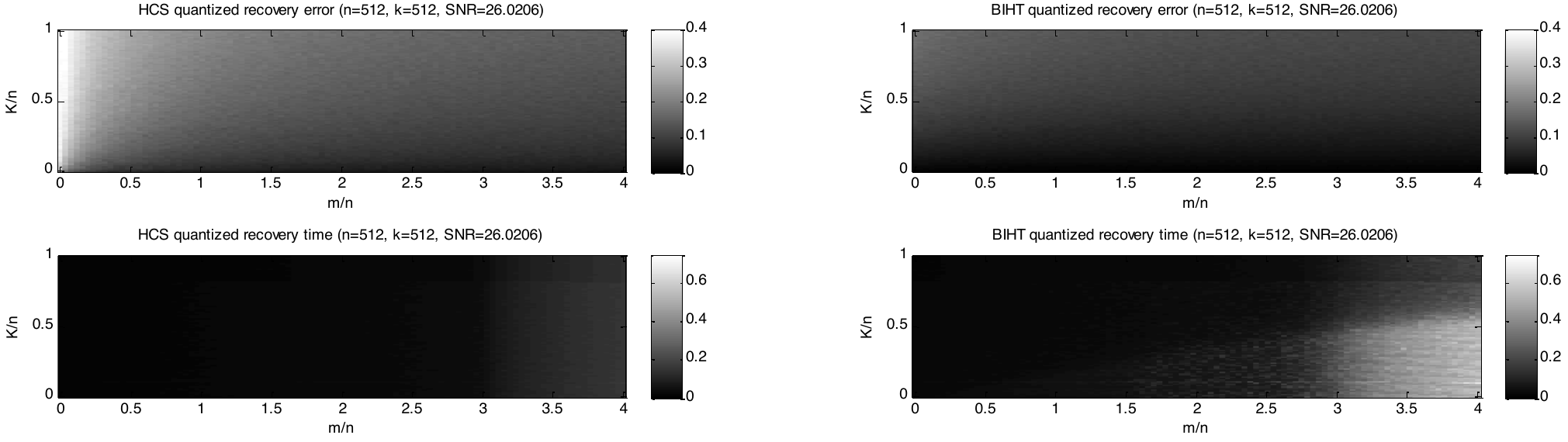}
}
\subfigure{
 \includegraphics[width=1\linewidth]{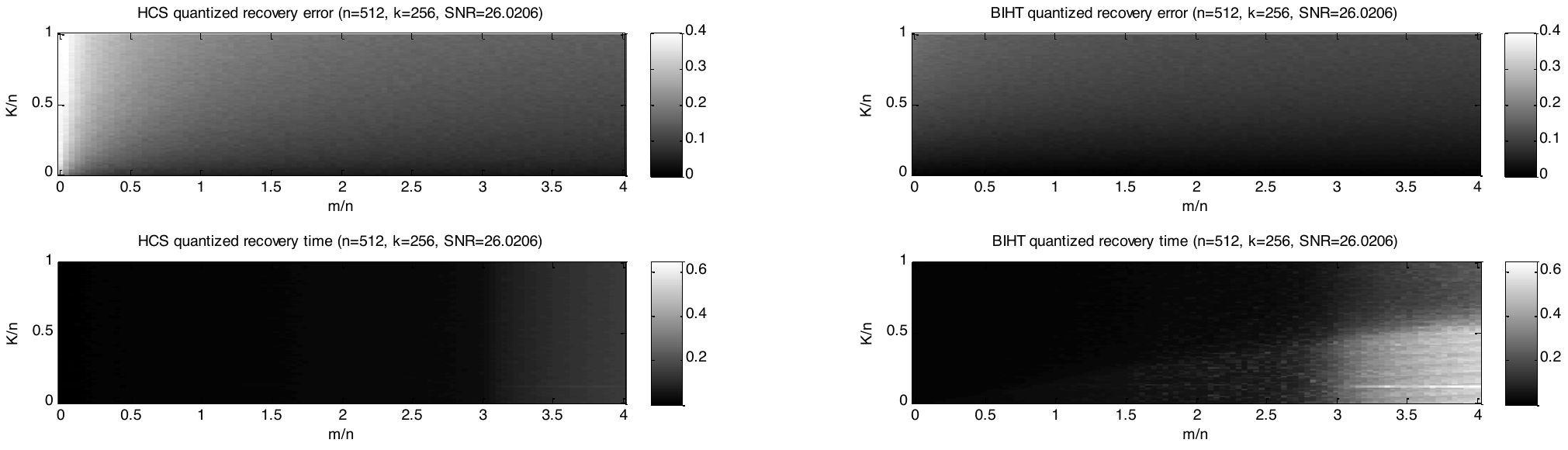}
}
\subfigure{
 \includegraphics[width=1\linewidth]{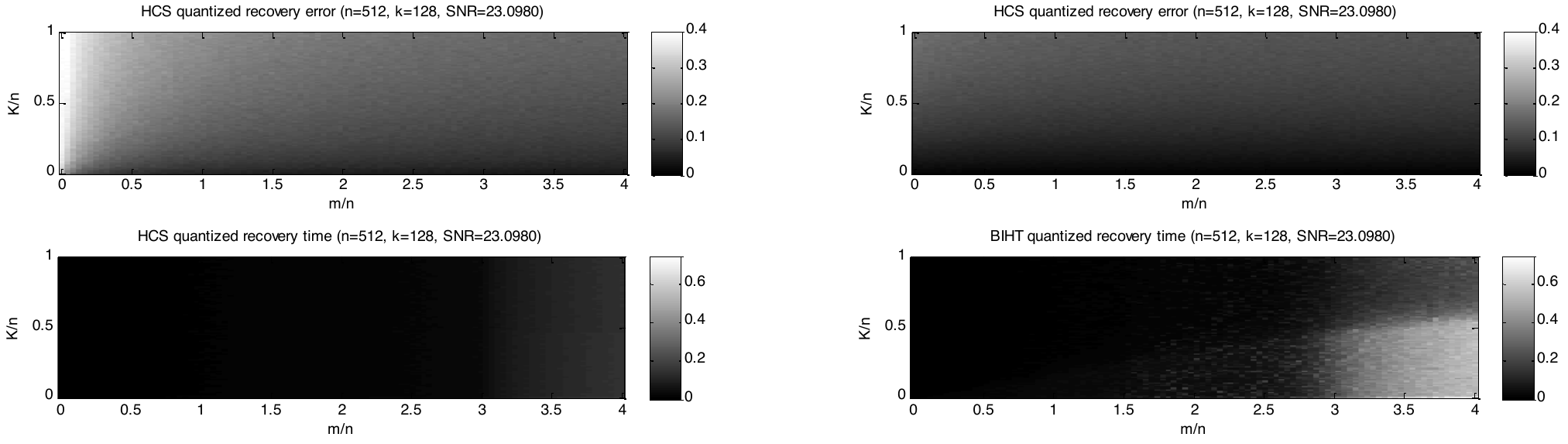}
}
\end{center}
 \caption{Phase plots of HCS and ``1-bit CS+HCS quantizer'' in the noisy case.}
\label{fig:nphase2}
\end{figure}

We also consider the phase transition properties \cite{PhaseT} of  HCS and 1-bit CS on quantized recovery error and on recovery time in the noisy case. The experiments setting up is the same as that in the noiseless case except Gaussian random noises are imposed to the input signals. The results are shown in Figure \ref{fig:nphase1} and Figure \ref{fig:nphase2}.

Comparing to the phase plots of quantized recovery error in the noiseless case, HCS performs much more robust than 1-bit CS. The time costs of HCS shown in Figure \ref{fig:nphase1} and Figure \ref{fig:nphase2} still significantly less than that of 1-bit CS.

\subsection{Quantized recovery error vs. number of measurements in the noisy case}

\begin{figure}[htb]
\begin{center}
\subfigure{
 \includegraphics[width=0.8\linewidth]{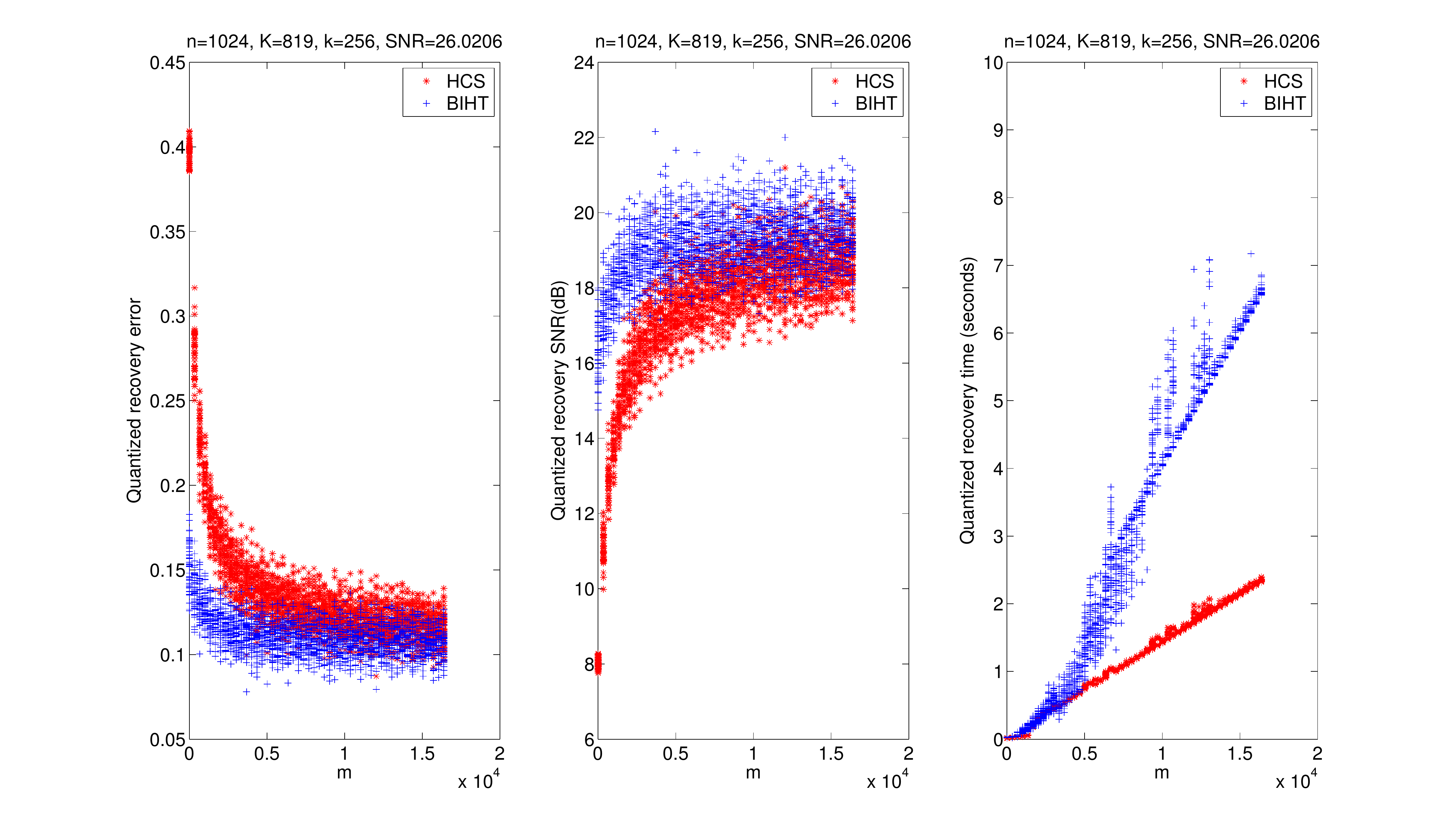}
}
\subfigure{
 \includegraphics[width=0.8\linewidth]{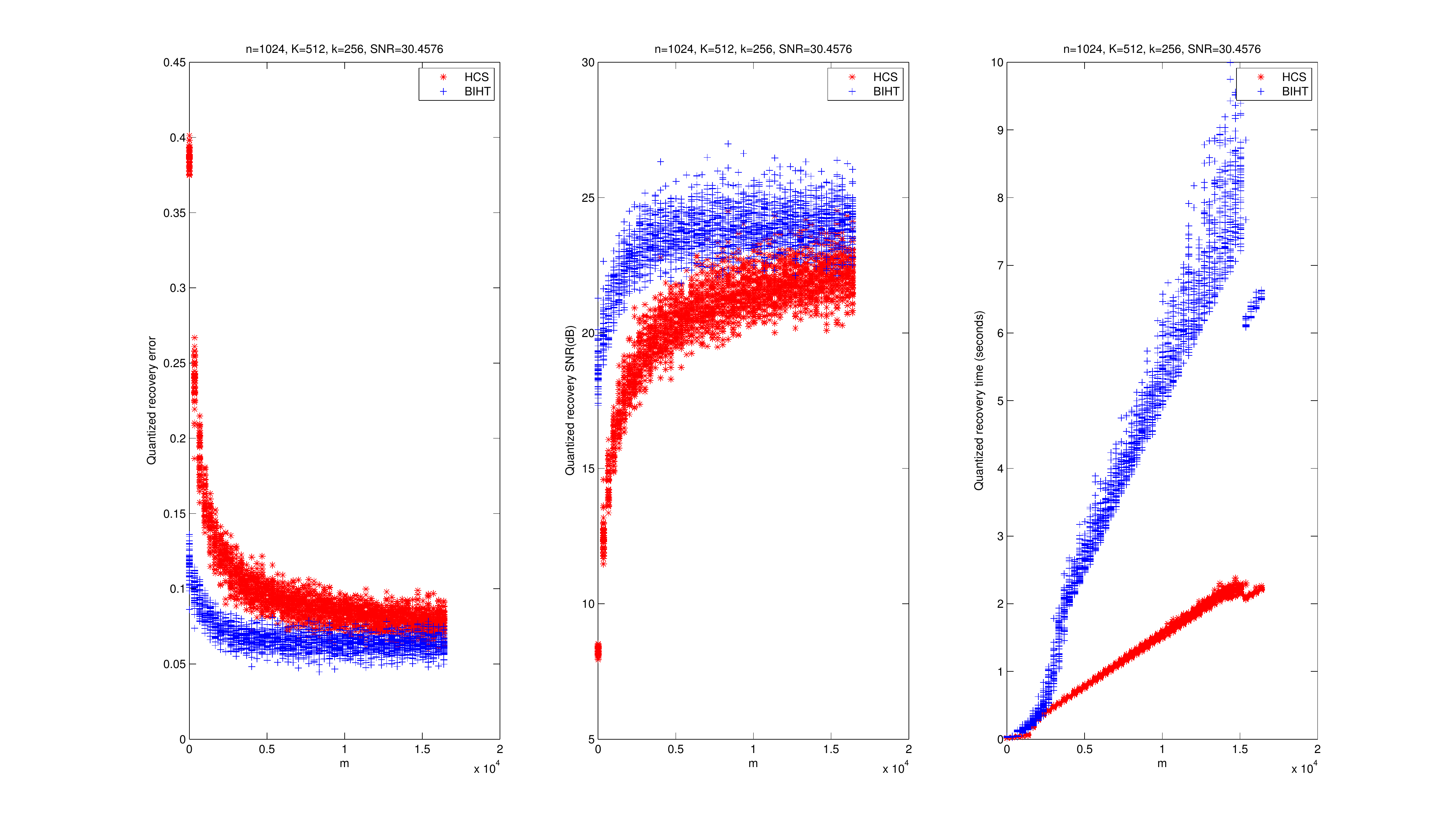}
}
\subfigure{
 \includegraphics[width=0.8\linewidth]{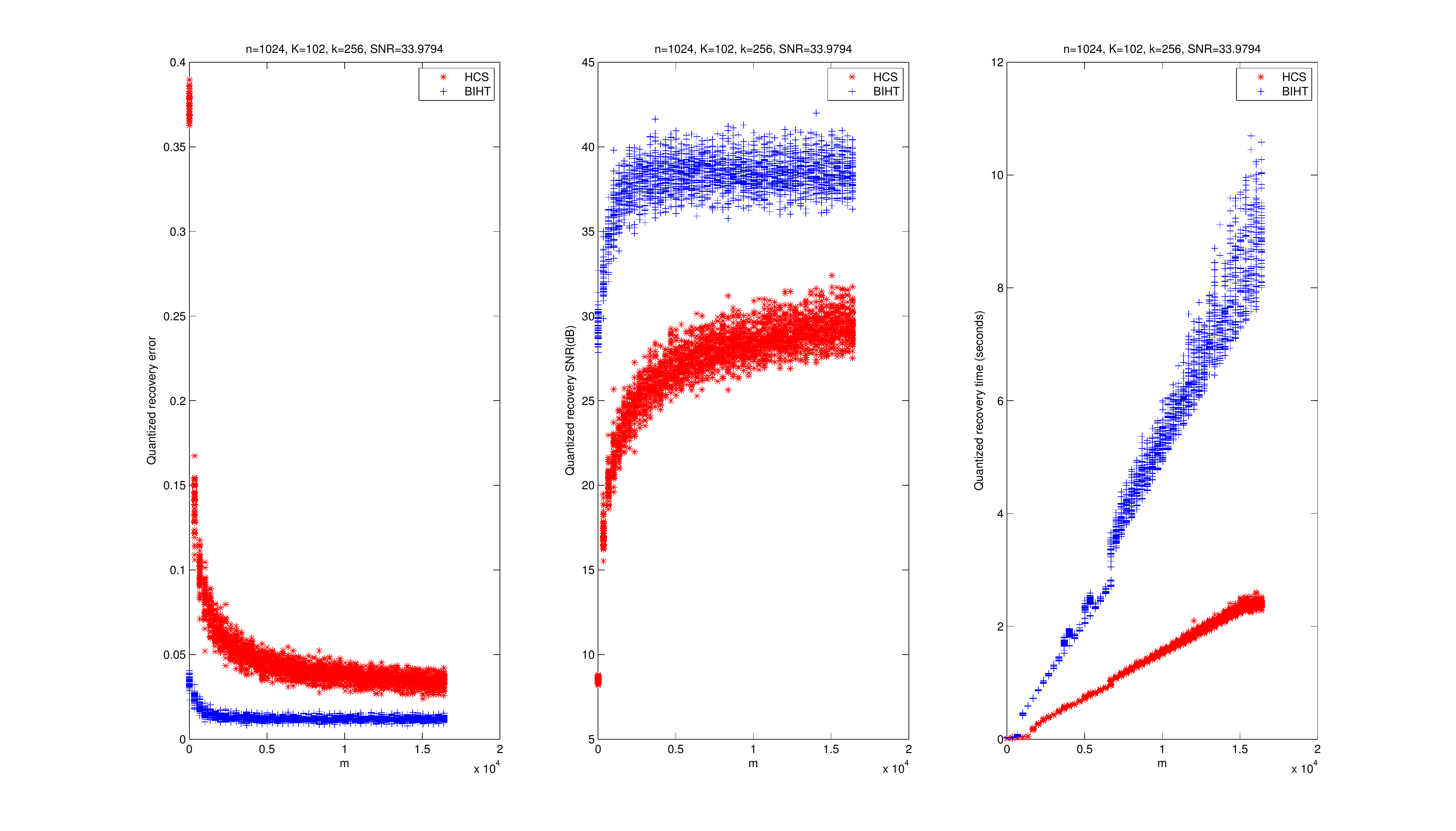}
}
\end{center}
 \caption{Quantized recovery error vs. number of measurements of HCS and ``1-bit CS+HCS quantizer'' in the noisy case.}
\label{fig:error1}
\end{figure}

\begin{figure}[htb]
\begin{center}
\subfigure{
 \includegraphics[width=0.8\linewidth]{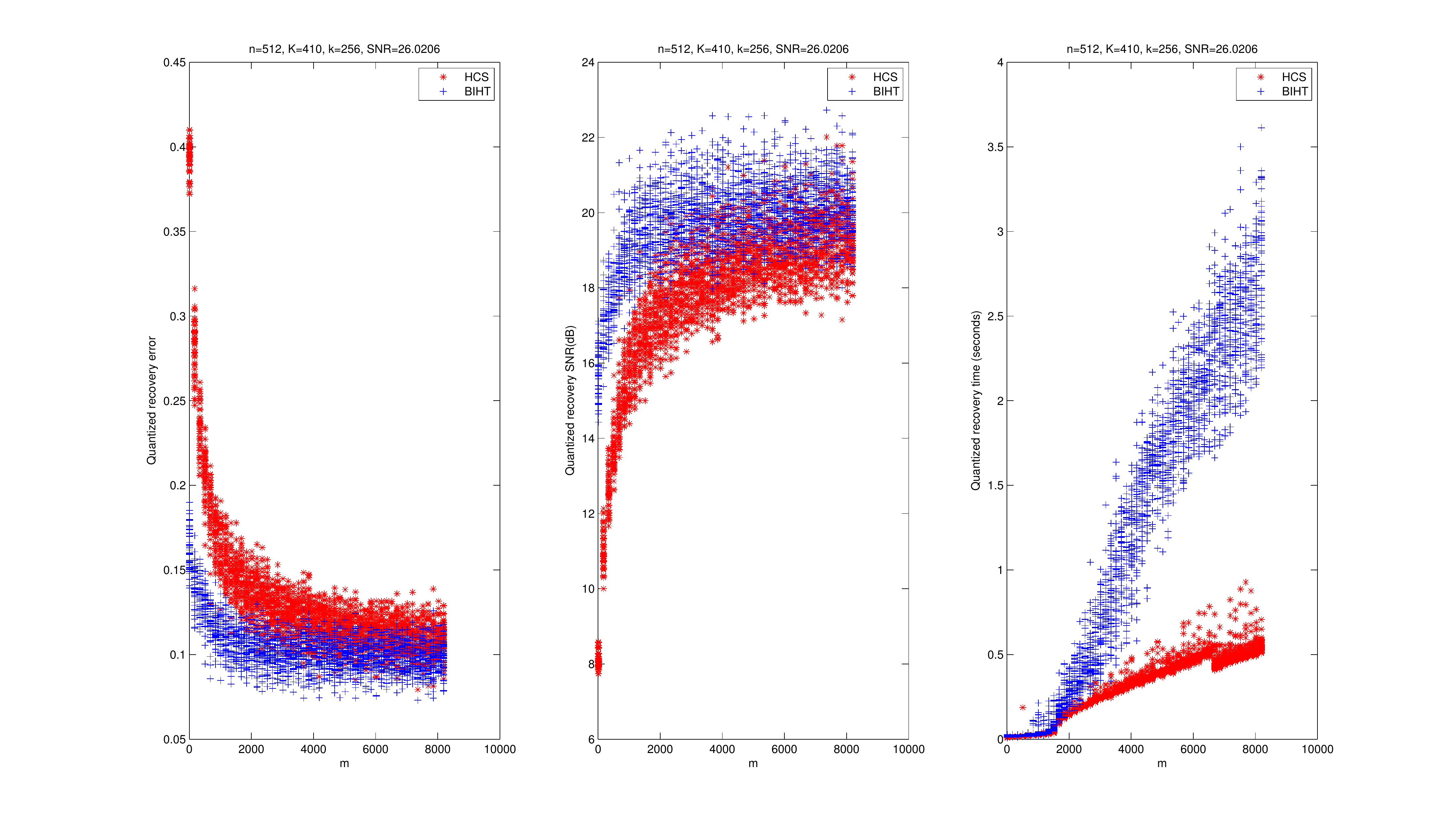}
}
\subfigure{
 \includegraphics[width=0.8\linewidth]{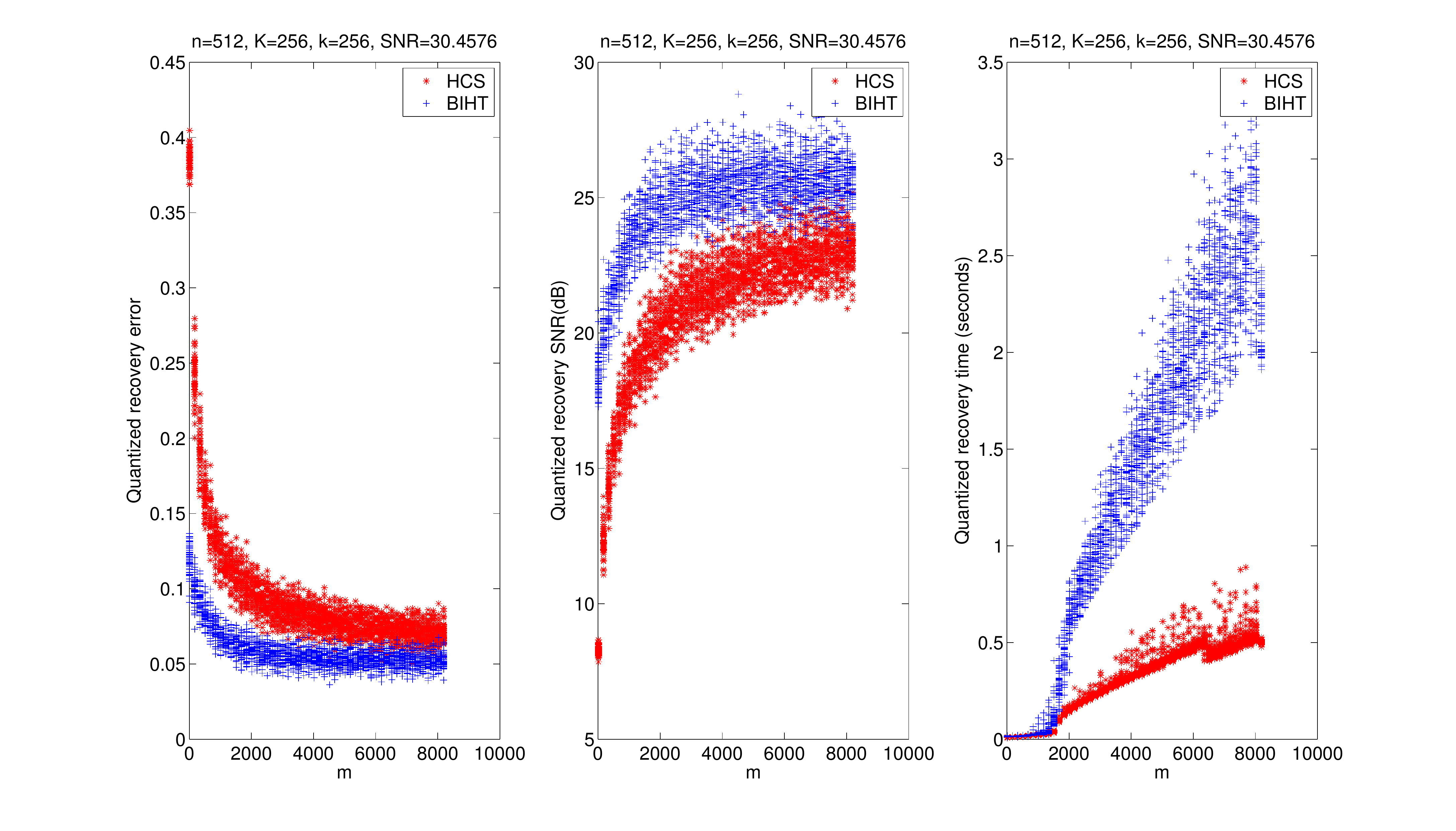}
}
\subfigure{
 \includegraphics[width=0.8\linewidth]{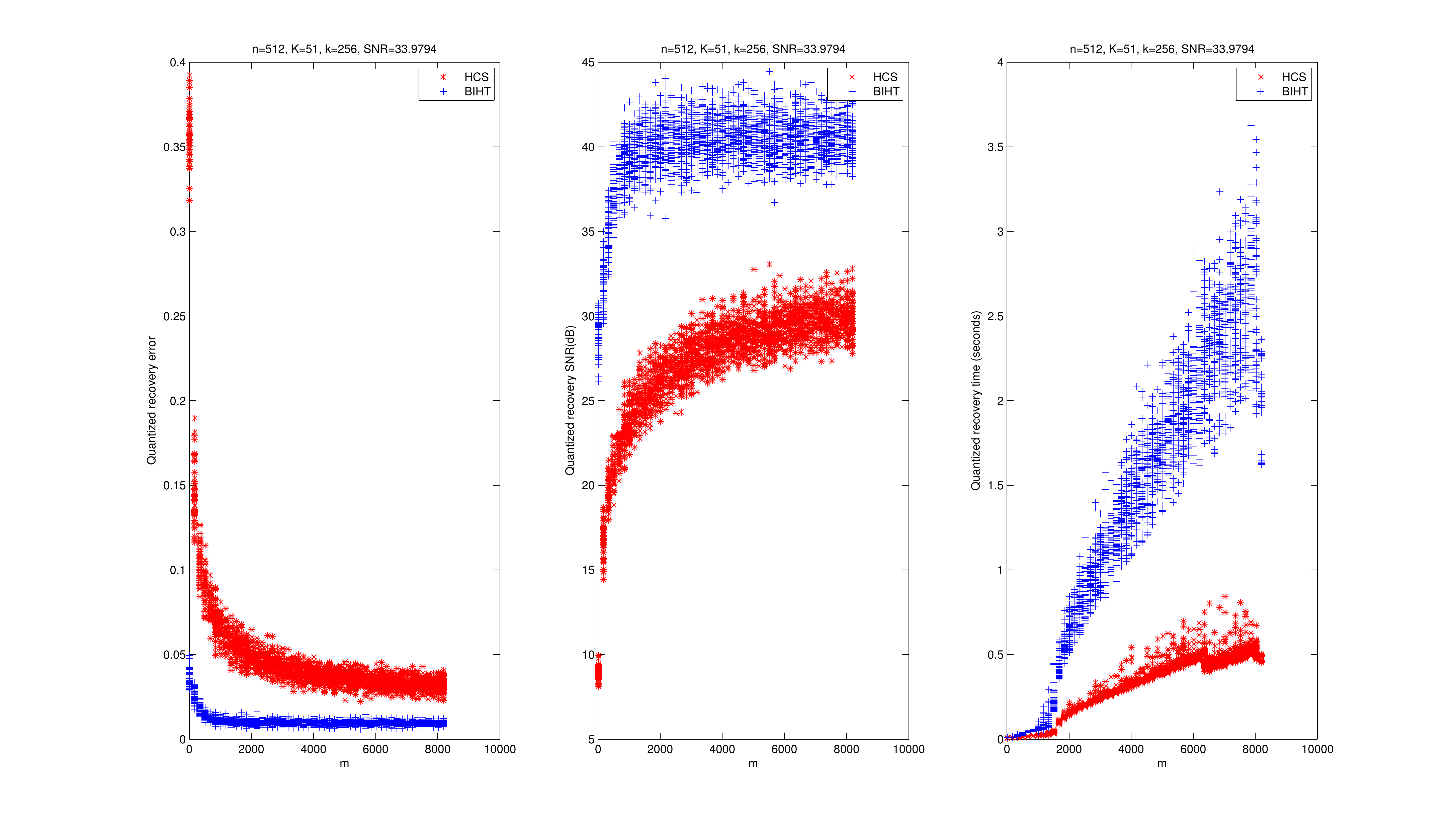}
}
\end{center}
 \caption{Quantized recovery error vs. number of measurements of HCS and ``1-bit CS+HCS quantizer'' in the noisy case.}
\label{fig:error2}
\end{figure}

We then show the trade-off between quantized recovery error and the amount of measurements on $2500$ trials for noisy signals of different $n$, $K$, $k$ and signal-to-noise ratio (SNR). In particular, given fixed $n$, $K$, $k$ and SNR, we uniformly choose $50$ values of $m$ between $0$ and $16n$. For each $m$ value, we conduct $50$ trials of HCS recovery and ``1-bit CS+HCS quantizer'' by recovering the quantizations of $50$ noisy signals from their $m$ 1-bit measurements. The quantized recovery error and time cost of each trial for different $n$, $K$, $k$ and SNR are shown in Figure \ref{fig:error1} and Figure \ref{fig:error2}.

Figure \ref{fig:error1} and Figure \ref{fig:error2} show the quantized recovery error of both HCS and ``1-bit CS+HCS quantization'' drops drastically with the increasing of the number of measurements. For dense signals with large noise, the two methods perform nearly the same on the recovery accuracy. This phenomenon indicates that HCS works well on dense signals and is robust to noise comparing to CS and 1-bit CS. In addition, the time cost of HCS increases substantially slower than that of ``1-bit CS+HCS quantizer'' with the increasing of the number of measurements.

\subsection{Dequantization and consistency}

\begin{figure}[htb]
\begin{center}
\subfigure{
 \includegraphics[width=1.05\linewidth]{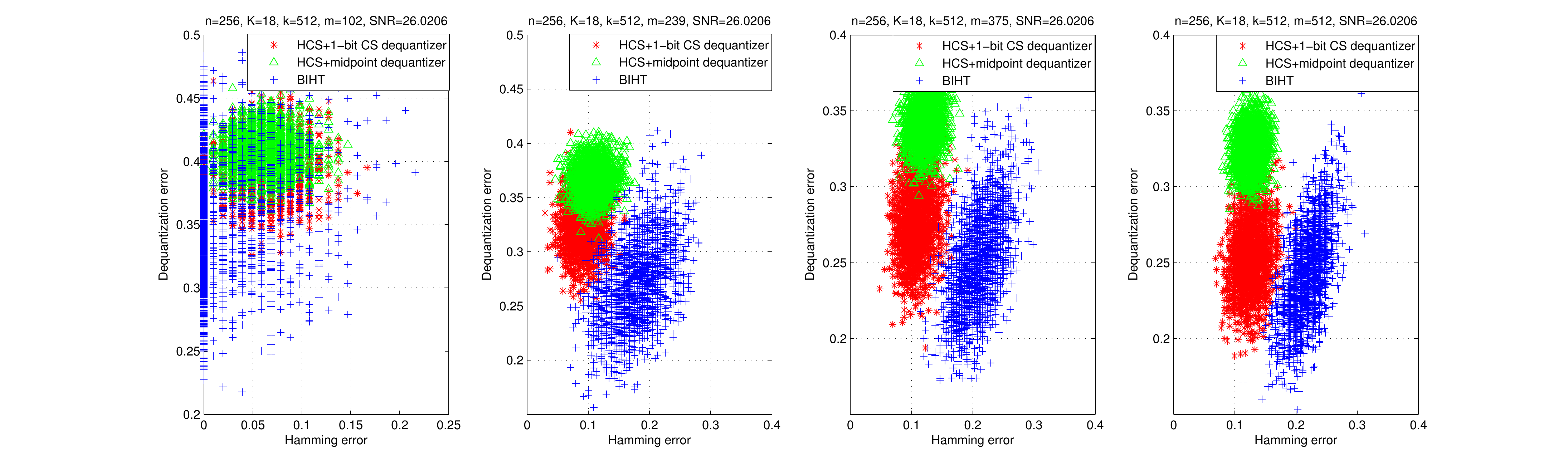}
}
\subfigure{
 \includegraphics[width=1.05\linewidth]{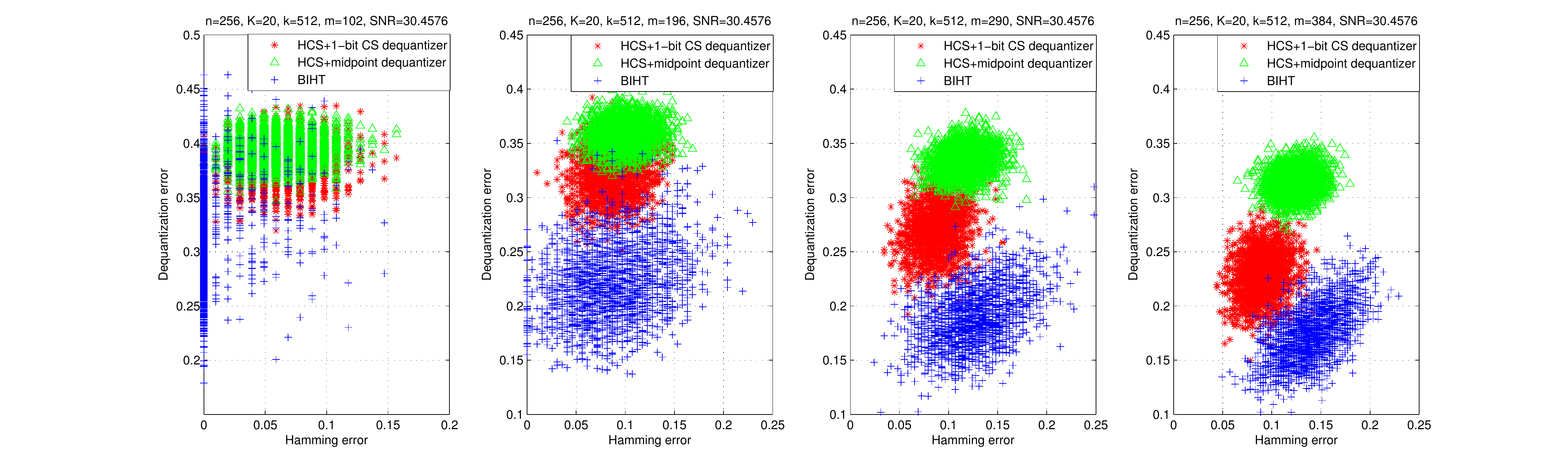}
}
\subfigure{
 \includegraphics[width=1.05\linewidth]{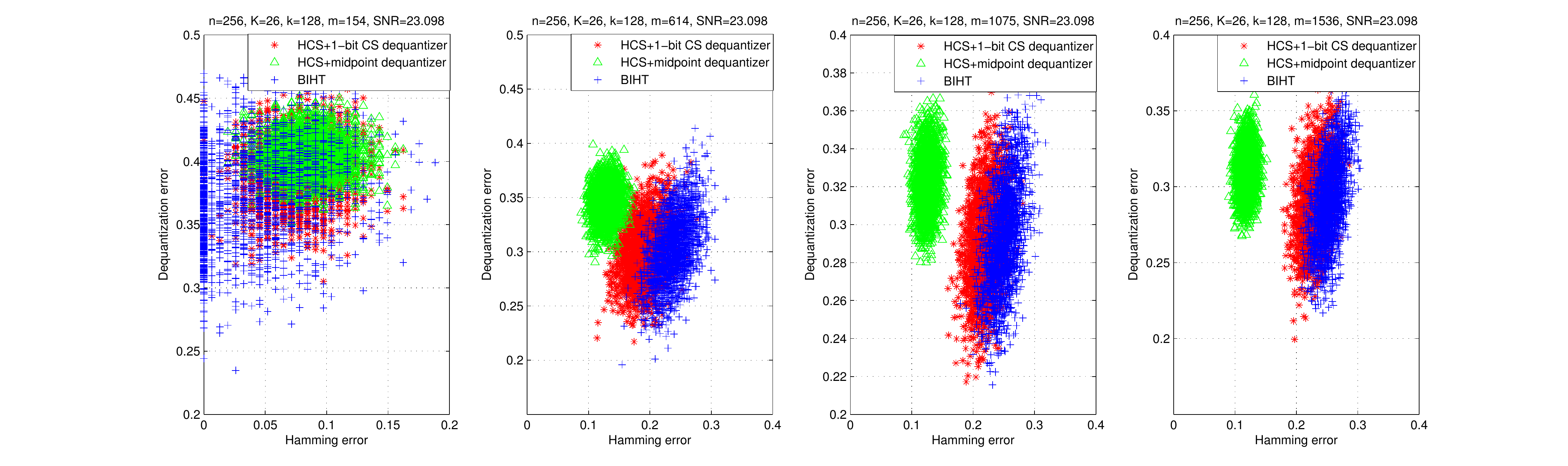}
}
\end{center}
 \caption{Reconstruction angular error vs. measurement Hamming error of ``HCS+1-bit CS dequantizer'', ``HCS+midpoint dequantizer'' and 1-bit CS in the noisy case.}
\label{fig:consis}
\end{figure}

We finally explore the performance of ``HCS+dequantizer'' stated in Section 4 and verify the consistency investigated in Lemma \ref{L:Consis}. In particular, we plot the normalized Hamming loss (defined in Lemma \ref{L:Consis}) between $A(x^*)$ and $A(x)$ vs. the angular error (\ref{E:BeSE}) between $x^*$ and $x$ of $2000$ trials for different amount of measurements in Figure \ref{fig:consis}.

Figure \ref{fig:consis} shows the linear relationship between Hamming error $D_H(A(x),A(x^*))$ and angular error $D_S(x,x^*)$ in ``HCS+1-bit CS dequantizer'', ``HCS+midpoint dequantizer'' and 1-bit CS given sufficient measurements. This linear relationship verifies the ``HCS+dequantizer'' error bound in Theorem \ref{T:Dbound} and the consistency in Lemma 1 of the Hamming Compressed Sensing submission. The figure also shows that ``HCS+1-bit CS dequantizer'' and 1-bit CS  perform better than ``HCS+midpoint dequantizer''. This verifies the effectiveness of ``1-bit CS dequantizer''. In the experiments, the `1-bit CS dequantizer'' only requires iterates less than $10$ to reach the accuracy obtained by 1-bit CS with more than $50$ iterates. Thus HCS can significantly save the computation of the subsequent dequantization.

\section{Conclusion}

We have proposed a new signal acquisition technique ``Hamming Compressed Sensing (HCS)'' to recover the k-bit quantization of a signal $x$ from a small amount of its 1-bit measurements. HCS recovery invokes $n$ times of KL-divergence based nearest neighbor searching in a Bernoulli distribution domain and requires only $nk$ computations of KL-divergence. The main significance of HCS is as follows: 1) it provides a direct recovery of quantized signal from a few measurements for digital systems, which has not been thoroughly studied but is essential in practice; 2) it has linear recovery time and thus its speed is extremely faster than optimization based or iterative methods; 3) the sparse assumption to signal is not compulsive in HCS. Another compelling advantage of HCS is that its recovery can significantly accelerate the subsequent dequantization. The quantized error bound of HCS for general signals and ``HCS+dequantizer'' recovery error bound for sparse signals have been carefully studied.

\ifCLASSOPTIONcaptionsoff
  \newpage
\fi



%
\bibliographystyle{IEEEtran}
\bibliography{HCS}

\begin{thebibliography}{10}
\providecommand{\url}[1]{#1}
\csname url@samestyle\endcsname
\providecommand{\newblock}{\relax}
\providecommand{\bibinfo}[2]{#2}
\providecommand{\BIBentrySTDinterwordspacing}{\spaceskip=0pt\relax}
\providecommand{\BIBentryALTinterwordstretchfactor}{4}
\providecommand{\BIBentryALTinterwordspacing}{\spaceskip=\fontdimen2\font plus
\BIBentryALTinterwordstretchfactor\fontdimen3\font minus
  \fontdimen4\font\relax}
\providecommand{\BIBforeignlanguage}[2]{{%
\expandafter\ifx\csname l@#1\endcsname\relax
\typeout{** WARNING: IEEEtran.bst: No hyphenation pattern has been}%
\typeout{** loaded for the language `#1'. Using the pattern for}%
\typeout{** the default language instead.}%
\else
\language=\csname l@#1\endcsname
\fi
#2}}
\providecommand{\BIBdecl}{\relax}
\BIBdecl

\bibitem{Shannon}
C.~E. Shannon, ``Communication in the presence of noise,'' \emph{Proceedings of
  Institute of Radio Engineers}, vol.~37, no.~1, pp. 10--21, 1949.

\bibitem{DonohoCS}
D.~L. Donoho, ``Compressed sensing,'' \emph{IEEE Transactions on Information
  Theory}, vol.~52, no.~4, pp. 1289--1306, 2006.

\bibitem{CandesT06}
E.~J. Cand{\`e}s and T.~Tao, ``Near-optimal signal recovery from random
  projections: Universal encoding strategies?'' \emph{IEEE Transactions on
  Information Theory}, vol.~52, no.~12, pp. 5406--5425, 2006.

\bibitem{ErrorCS}
E.~J. Cand\`{e}s, M.~Rudelson, T.~Tao, and R.~Vershynin, ``Error correction via
  linear programming,'' \emph{Foundations of Computer Science, Annual IEEE
  Symposium on}, pp. 295--308, 2005.

\bibitem{ModelCS}
R.~G. Baraniuk, V.~Cevher, M.~F. Duarte, and C.~Hegde, ``Model-based
  compressive sensing,'' \emph{IEEE Transactions on Information Theory},
  vol.~56, pp. 1982--2001, 2010.

\bibitem{BayesianCS}
S.~Ji, Y.~Xue, and L.~Carin, ``Bayesian compressive sensing,'' \emph{IEEE
  Transactions on Signal Processing}, vol.~56, no.~6, pp. 2346--2356, 2008.

\bibitem{SparseCS}
A.~Gilbert and P.~Indyk, ``Sparse recovery using sparse matrices,''
  \emph{Proceedings of the IEEE}, vol.~98, no.~6, pp. 937--947, 2010.

\bibitem{OMP}
J.~A. Tropp and A.~C. Gilbert, ``Signal recovery from random measurements via
  orthogonal matching pursuit,'' \emph{IEEE Transactions on Information
  Theory}, vol.~53, pp. 4655--4666, 2007.

\bibitem{CoSaMP}
D.~Needell and J.~A. Tropp, ``Cosamp: Iterative signal recovery from incomplete
  and inaccurate samples,'' \emph{Applied and Computational Harmonic Analysis},
  vol.~26, pp. 301--321, 2008.

\bibitem{MessagePassing}
D.~L. Donoho, A.~Maleki, and A.~Montanari, ``Message passing algorithms for
  compressed sensing,'' \emph{Proceedings of the National Academy of Sciences},
  2009.

\bibitem{IST}
I.~Daubechies, M.~Defrise, and C.~D. Mol, ``An iterative thresholding algorithm
  for linear inverse problems with a sparsity constraint,''
  \emph{Communications on Pure and Applied Mathematics}, vol.~57, no.~11, pp.
  1413--1457, 2004.

\bibitem{IterativeHardThresh}
K.~Bredies and D.~A. Lorenz, ``Iterated hard shrinkage for minimization
  problems with sparsity constraints,'' \emph{SIAM Journal on Scientific
  Computing}, vol.~30, no.~2, pp. 657--683, 2008.

\bibitem{BasisPursuit}
S.~S. Chen, D.~L. Donoho, and M.~A. Saunders, ``Atomic decomposition by basis
  pursuit,'' \emph{SIAM Journal on Scientific Computing}, vol.~20, no.~1, pp.
  33--61, 1999.

\bibitem{DantzigSelector}
E.~J. Cand{\`e}s and T.~Tao, ``The dantzig selector: statistical estimation
  when p is much larger than n,'' \emph{Annals of Statistics}, vol.~35, no.~6,
  pp. 2313--2351, 2007.

\bibitem{NESTA}
J.~Bobin, S.~Becker, and E.~Cand{\`e}s, ``Nesta: A fast and accurate
  first-order method for sparse recovery,'' \emph{technical report}, 2009.

\bibitem{InteriorPoint}
S.-J. Kim, K.~Koh, M.~Lustig, S.~Boyd, and D.~Gorinevsky, ``An interior-point
  method for large-scale l1-regularized least squares,'' \emph{IEEE Journal of
  In Selected Topics in Signal Processing}, vol.~1, no.~4, pp. 606--617, 2007.

\bibitem{CoordinateDescent}
P.~Tseng and S.~Yun, ``A coordinate gradient descent method for nonsmooth
  separable minimization,'' \emph{Mathematical Programming}, vol. 117, no. 1-2,
  pp. 387--423, 2009.

\bibitem{GPSR}
M.~A.~T. Figueiredo, R.~D. Nowak, and S.~J. Wright, ``Gradient projection for
  sparse reconstruction: Application to compressed sensing and other inverse
  problems,'' \emph{IEEE Journal of Selected Topics in Signal Processing},
  vol.~1, no.~4, pp. 586--597, 2007.

\bibitem{BregmanIteration}
W.~Yin, S.~Osher, D.~Goldfarb, and J.~Darbon, ``Bregman iterative algorithms
  for l1-minimization with applications to compressed sensing,'' \emph{SIAM
  Journal on Imaging Sciences}, vol.~1, no.~1, pp. 143--168, 2008.

\bibitem{FPC}
E.~T. Hale, W.~Yin, and Y.~Zhang, ``Fixed-point continuation for
  $\ell_1$-minimization: Methodology and convergence,'' \emph{SIAM Journal on
  Optimization}, vol.~19, no.~3, pp. 1107--1130, 2008.

\bibitem{IRLS}
B.~Rao, K.~Engan, S.~Cotter, J.~Palmer, and K.~Kreutz-Delgado, ``Subset
  selection in noise based on diversity measure minimization,'' \emph{IEEE
  Transactions on Signal Processing}, vol.~51, pp. 760--770, 2003.

\bibitem{lasso}
R.~Tibshirani, ``Regression shrinkage and selection via the lasso,''
  \emph{Journal of the Royal Statistical Society (Series B)}, vol.~58, pp.
  267--288, 1996.

\bibitem{LARS}
B.~Efron, T.~Hastie, I.~Johnstone, and R.~Tibshirani, ``Least angle
  regression,'' \emph{Annals of Statistics}, vol.~32, no.~2, pp. 407--499,
  2004.

\bibitem{GroupLasso}
M.~Yuan and Y.~Lin, ``Model selection and estimation in regression with grouped
  variables,'' \emph{Journal of the Royal Statistical Society (Series B)},
  vol.~68, pp. 49--67, 2006.

\bibitem{CandesRT06}
E.~J. Cand{\`e}s, J.~K. Romberg, and T.~Tao, ``Robust uncertainty principles:
  exact signal reconstruction from highly incomplete frequency information,''
  \emph{IEEE Transactions on Information Theory}, vol.~52, no.~2, pp. 489--509,
  2006.

\bibitem{StableCS}
E.~J. Cand\`{e}s, J.~K. Romberg, and T.~Tao, ``{Stable signal recovery from
  incomplete and inaccurate measurements},'' \emph{Communications on Pure and
  Applied Mathematics}, vol.~59, no.~8, pp. 1207--1223, 2006.

\bibitem{EldarRobust}
Y.~C. Eldar and M.~Mishali, ``Robust recovery of signals from a structured
  union of subspaces,'' \emph{IEEE Transactions on Information Theory},
  vol.~55, pp. 5302--5316, 2009.

\bibitem{CSreview}
A.~M. Bruckstein, D.~L. Donoho, and M.~Elad, ``From sparse solutions of systems
  of equations to sparse modeling of signals and images,'' \emph{SIAM Review},
  vol.~51, no.~1, pp. 34--81, 2009.

\bibitem{BPDNQ}
L.~Jacques, D.~K. Hammond, and M.-J. Fadili, ``Dequantizing compressed sensing:
  When oversampling and non-gaussian constraints combine,'' \emph{IEEE
  Transactions on Information Theory}, vol.~57, no.~1, pp. 559--571, 2011.

\bibitem{QCSboyd}
A.~Zymnis, S.~Boyd, and E.~Cand\`{e}s, ``Compressed sensing with quantized
  measurements,'' \emph{IEEE Signal Processing Letters}, vol.~17, no.~2, pp.
  149--152, 2010.

\bibitem{1bitCS}
P.~T. Boufounos and R.~G. Baraniuk, ``One-bit compressive sensing,'' in
  \emph{Conference on Information Sciences and Systems (CISS)}, 2008.

\bibitem{Distorsion}
W.~Dai, H.~V. Pham, and O.~Milenkovic, ``Distortion-rate functions for
  quantized compressive sensing,'' in \emph{IEEE Information Theory Workshop
  (ITW)}, 2009, pp. 171--175.

\bibitem{SubspacePursuit}
W.~Dai and O.~Milenkovic, ``Subspace pursuit for compressive sensing signal
  reconstruction,'' \emph{IEEE Transactions on Information Theory}, vol.~55,
  no.~5, pp. 2230--2249, 2009.

\bibitem{Robust1bitCS}
L.~Jacques, J.~N. Laska, P.~T. Boufounos, and R.~G. Baraniuk, ``Robust 1-bit
  compressive sensing via binary stable embeddings of sparse vectors,''
  \emph{arXiv:1104.3160}, 2011.

\bibitem{Greedy1bitCS}
P.~T. Boufounos, ``Greedy sparse signal reconstruction from sign
  measurements,'' in \emph{Proc. Asilomar Conference on Signals Systems and
  Computers}, 2009.

\bibitem{Trust}
J.~N. Laska, Z.~Wen, W.~Yin, and R.~G. Baraniuk, ``Trust, but verify: Fast and
  accurate signal recovery from 1-bit compressive measurements,'' \emph{Rice
  University CAAM Technical Report TR10-30}, 2010.

\bibitem{LSH}
P.~Indyk and R.~Motwani, ``Approximate nearest neighbors: towards removing the
  curse of dimensionality,'' in \emph{Proceedings of the thirtieth annual ACM
  symposium on Theory of computing (STOC)}, 1998, pp. 604--613.

\bibitem{LSH2008}
A.~Andoni and P.~Indyk, ``Near-optimal hashing algorithms for approximate
  nearest neighbor in high dimensions,'' \emph{Communications of the ACM},
  vol.~51, pp. 117--122, 2008.

\bibitem{AppMaxCut}
M.~X. Goemans and D.~P. Williamson, ``Improved approximation algorithms for
  maximum cut and satisfiability problems using semidefinite programming,''
  \emph{Journal of the ACM}, vol.~42, no.~6, pp. 1115--1145, 1995.

\bibitem{PhaseT}
D.~L. Donoho, A.~Maleki, and A.~Montanari, ``The noise-sensitivity phase
  transition in compressed sensing,'' \emph{arXiv:1004.1218}, 2010.

\end{thebibliography}

%

%





\end{document}